\documentclass[%
 reprint,
nofootinbib,
 amsmath,amssymb,
 aps,
floatfix,
]{revtex4-2}

\usepackage{graphicx}
\usepackage{dcolumn}
\usepackage{bm}
\usepackage{txfonts}
\usepackage{natbib}
\usepackage{xcolor}
\usepackage{subcaption}
\usepackage[hyperindex,breaklinks=true,colorlinks,citecolor=blue]{hyperref} 
\usepackage{stackengine}
\usepackage{algorithmic}
\usepackage{booktabs}       
\usepackage{amsfonts}       
\usepackage{nicefrac}       
\usepackage{microtype}      
\usepackage{txfonts}

\newcommand\xrowht[2][0]{\addstackgap[.5\dimexpr#2\relax]{\vphantom{#1}}}

\newcommand{\dd}{\mbox{d}}

\newcommand\quedalle[1]{}

\begin{document}

\preprint{APS/123-QED}

\title{New interpretable statistics for large-scale structure analysis and generation}

\author{E. Allys}
\thanks{Both authors contributed equally to this work.}
\affiliation{%
Laboratoire de Physique de l'\'Ecole Normale Sup\'erieure, ENS, Universit\'e PSL, CNRS, Sorbonne Universit\'e, Universit\'e de Paris, F-75005 Paris, France
}%
\author{\vspace{-0.3cm}T. Marchand}%
\thanks{Both authors contributed equally to this work.}
\affiliation{%
DI, \'Ecole Normale Sup\'erieure, ENS, Universit\'e PSL, Paris, France \\
Laboratoire de Physique de l'\'Ecole Normale Sup\'erieure, ENS, Universit\'e PSL, CNRS, Sorbonne Universit\'e, Universit\'e de Paris, F-75005 Paris, France\\
}%
\author{\vspace{-0.6cm}J.-F. Cardoso}
\affiliation{
CNRS and Sorbonne Universit\'e, UMR 7095, Institut d'Astrophysique de Paris, 98 bis Boulevard Arago, 75014 Paris, France
}%
\author{\vspace{-0.2cm}F. Villaescusa-Navarro}
\affiliation{%
Department of Astrophysical Sciences, Princeton University, Peyton Hall, Princeton NJ 08544, USA \\
Center for Computational Astrophysics, Flatiron Institute, 162 5th Avenue, New York NY 10010, USA
}%
\author{\vspace{-0.3cm}S. Ho}
\affiliation{%
Center for Computational Astrophysics, Flatiron Institute, 162 5th Avenue, New York NY 10010, USA \\
Department of Astrophysical Sciences, Princeton University, Peyton Hall, Princeton NJ 08544, USA
}%
\author{\vspace{-0.2cm}S. Mallat}
\affiliation{%
Coll\`ege de France, Paris, France \\
Center for Computational Mathematics, Flatiron Institute, 162 5th Avenue, New York NY 10010, USA\\
DI, \'Ecole Normale Sup\'erieure, ENS, Universit\'e PSL, Paris, France
}%

\date{\today}

\begin{abstract}
We introduce Wavelet Phase Harmonics (WPH) statistics: interpretable low-dimensional statistics that describe 2D non-Gaussian fields. These statistics are built from WPH moments, which were recently introduced in the data science and machine learning community. We apply WPH statistics to projected 2D matter density fields from the Quijote N-body simulations of the large-scale structure of the Universe. 
By computing Fisher information matrices, we find that the WPH statistics place more stringent constraints on four of five cosmological parameters when compared to statistics based on the combination of the power spectrum and bispectrum. We also use the WPH statistics with a maximum entropy model to statistically generate new 2D density fields that accurately reproduce the probability density function, the mean and standard deviation of the power spectrum, the bispectrum, and Minkowski functionals of the input density fields. 
Although other methods are efficient for either parameter estimates or statistical syntheses of the large-scale structure, WPH statistics are the first statistics that achieve state-of-the-art results for both tasks as well as being interpretable.
\end{abstract}

\maketitle


\section{Introduction}


The evolution of the large-scale structure (LSS) of the Universe
illustrates how nonlinearities can affect the statistical properties of a field.  The fluctuations of the density field are Gaussian in the early Universe, and then grow into a complex structure containing walls, filaments, nodes, and voids---the cosmic web. These structures are direct signatures of the coupling of the different scales in the cosmic web.

No generic and efficient statistical characterization of the LSS exists, in contrast to the cosmic microwave background (CMB) for which we have statistically extracted most of the information on the cosmological model. Indeed, we can characterize the primary temperature anisotropies of the CMB by a Gaussian field, and describe them fully by their power spectrum. In other words, for a homogeneous and isotropic Gaussian field such as the CMB, there is no interaction between different scales, and the amplitudes of its Fourier modes entirely characterize the field. 
Conversely, the LSS field is a non-Gaussian field with long-range interactions. The power spectrum alone cannot describe the couplings between different scales of the LSS.

A standard method to capture the nonlinearity of the LSS is to compute $n$-point correlation functions, which correspond to poly-spectra when expressed in terms of Fourier modes. In particular, various studies in recent decades rely on the bispectrum (poly-spectrum for $n=3$)  to study the LSS (e.g., see~\citep{Sefusatti:2006pa,Byun:2017fkz,hahn2019constraining}). 
One difficulty of directly using the Fourier bispectrum is its large number of terms, which generally must be reduced in some way. This typically leads to the construction of tailored bispectrum estimators (e.g., see~\citep{Chiang:2014oga}). 
In addition, bispectrum estimators, as with any high-order moments, are very sensitive to outliers and thus may suffer high empirical variance~\citep{stuart1963advanced}.

Alternatively, other studies have developed new statistics to go beyond bispectrum analysis of the LSS fields. For example, ~\citep{Obreschkow:2012yb, Wolstenhulme:2014cla, Alpaslan:2014ura} use the line correlation function (LCF) to characterize the LSS and to perform cosmological parameter inference. The LCF computes pure phase information in Fourier space and is particularly efficient for describing filamentary structures, especially when used in addition to the power spectrum and bispectrum~\citep{Byun:2017fkz,Ali:2018sdk}. There is also an abundant literature on other statistics, such as the distribution of peaks~\citep{bardeen1985statistics} or of voids~\citep{pisani2019cosmic} in the cosmic web. 

Non-Gaussian fields such as the LSS contain coherent structures at different scales that are well localized in space and in frequency. This feature motivates a hierarchical multiscale approach, such as the wavelet transform, rather than a description in terms of Fourier modes, which are not localized in space. The wavelet transform decomposes a process at different scales and locations and often leads to a sparse spatial description~\citep{cohen1995wavelets,mallat1999wavelet,van2004wavelets,farge2010multiscale,farge2015wavelet}.

However, the wavelet transform in itself does not characterize interactions between scales. Indeed, second-order moments of a wavelet transform depend solely on the power spectrum~\citep{flandrin1992wavelet,meyer1999wavelets,farge2010multiscale}. %
To capture the interactions between scales, we have to compute correlations between nonlinear transforms of the wavelet coefficients. This approach leads to statistical descriptors characterizing the dependences across different scales that are signatures of the coherent structures of the field.

Recently, \cite{mallat2018phase} introduced a novel low-dimensional statistical description following these principles called \emph{Wavelet Phase Harmonics} (WPH) statistics. The authors applied a nonlinear operator, the \textit{phase harmonic operator}, to the multiscale wavelet transform of a field. This operator acts on the complex phase of a field independently of the amplitude and enables alignment of the phase information across different scales. The building block of WPH statistics are WPH moments, i.e., covariances of wavelet transforms whose spatial frequencies have been made synchronous by means of the phase harmonic operator. WPH statistics are able to capture coupling between scales and can efficiently reproduce various textures~\citep{zhang2019maximum}. Moreover, they achieve competitive classification results on data sets as challenging as ImageNet~\citep{zarka2019deep}. 

Building upon these recent results, we design in this paper low-dimensional WPH statistics suited to the matter density field of the LSS\footnote{This work was done simultaneously and independently of that presented in \cite{cheng2020new}, where the authors apply a different but related technique, the wavelet scattering transform, to perform cosmological parameter inference in the context of weak lensing.}.
At present, we work with a 2D projection of the LSS matter density field.
We validate our newly constructed statistics by applying them to two complementary tasks: i) measuring cosmological information and ii) generating statistical syntheses. For the first task, we compute the Fisher information contained in these statistics with respect to five cosmological parameters. For the second task, we generate statistical syntheses of the 2D projected LSS matter density field by building a maximum-entropy generative model.  
Such a model generates new realizations of the field that are conditioned on the WPH statistics, while being \emph{as general as possible}. That is, the new realizations include no additional implicit or explicit constraints. We assess the quality of the syntheses by checking how well they reproduce standard cosmological statistics such as the power spectrum, bispectra, and Minkoswki functionals.

We obtain state-of-the-art results for both these tasks, which is the main result of this paper. Although previous approaches have been successful for one or the other of these tasks, to the best of our knowledge this is the first time that use of a single low-dimensional statistical description has achieved such performance on both of them.

Additionally, we demonstrate the interpretability of WPH statistics: they provide  better physical insight into the structure of the LSS matter density field. In particular, we see which features of the LSS are related to interactions between near and distant scales. We also discuss the relative impact on the different cosmological parameters of the coupling between different scales.

\begin{figure}[t]
\begin{subfigure}{.23\textwidth}
  \centering
  \includegraphics[width=\linewidth]{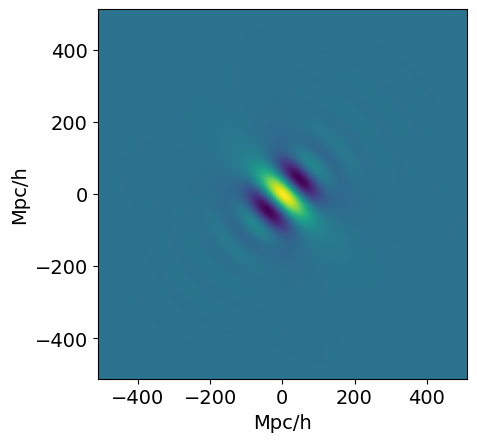}  
  \label{fig:wavelet}
\end{subfigure}
\begin{subfigure}{.23\textwidth}
  \centering
  \includegraphics[width=\linewidth]{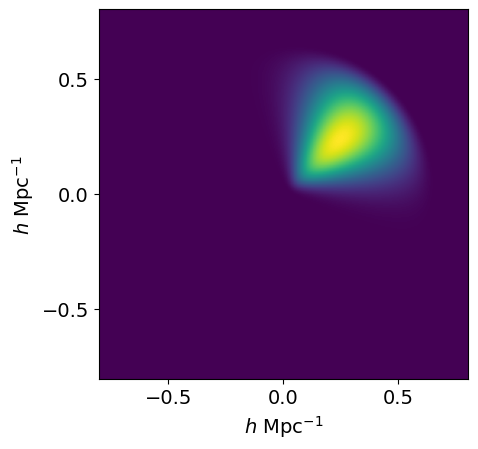}  
  \label{fig:fftwavelet}
  \end{subfigure}
  \vspace{-0.55cm}
  \caption{Two-dimensional bump steerable wavelets. The real part of $\psi_{4,2}(\vec{x})$ (left) and the Fourier transform $\hat{\psi}_{1,2}(\vec{k})$ (right). Axes are labeled with the units used when applying these wavelets to density maps of the Large Scale Structure (LSS).}
\label{FigWavelet2D}
\end{figure}

\paragraph*{Outline of the paper.}
We base our work on two-dimensional projected matter density fields from the Quijote N-body simulations of the LSS~\citep{villaescusa2019quijote}. We present in Sec.~\ref{PartWaveletToWPH} the general form of the low-dimensional WPH statistical description that we use throughout the paper. In Sec.~\ref{PartCosmoInfo}, we briefly describe the Quijote simulations, and present Fisher analysis results for five cosmological parameters based on the fields from these simulations. In Sec.~\ref{PartSyntheses}, we present the microcanonical maximum entropy generative model that we use, and we assess the quality of the statistical syntheses generated from WPH statistical constraints. Finally, we discuss in Sec.~\ref{PartDiscussionVariousTerms} the physical interpretation of the WPH coefficients, and their link with standard summary statistics. Appendix~\ref{AppMathSpec} specifies some mathematical details, including the form of the mother wavelet and the bispectrum statistics. Appendix~\ref{AppendixFinalModels} specifies the complete parameters of the WPH representations used to perform the cosmological Fisher analysis and statistical syntheses.

\begin{figure*}[t]
\begin{subfigure}{.32\textwidth}
  \centering
  \includegraphics[width=\linewidth]{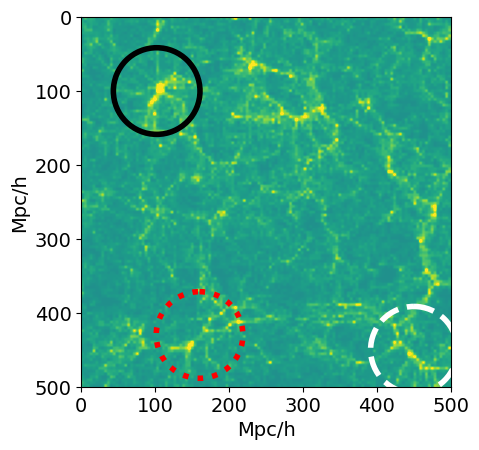}  
  \label{fig:QuijoteUnfiltered}
\end{subfigure}
\begin{subfigure}{.32\textwidth}
  \centering
  \includegraphics[width=\linewidth]{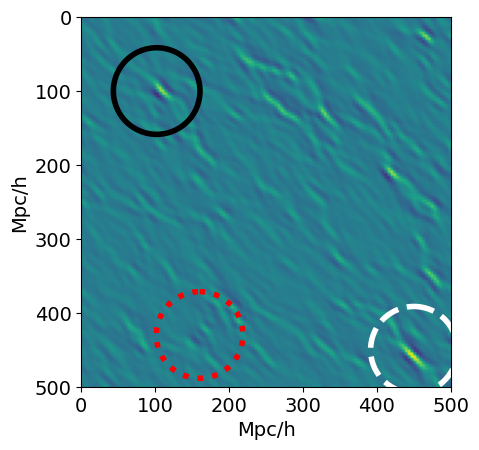} 
  \label{fig:QuijoteFiltered1}
\end{subfigure}
\begin{subfigure}{.32\textwidth}
  \centering
  \includegraphics[width=\linewidth]{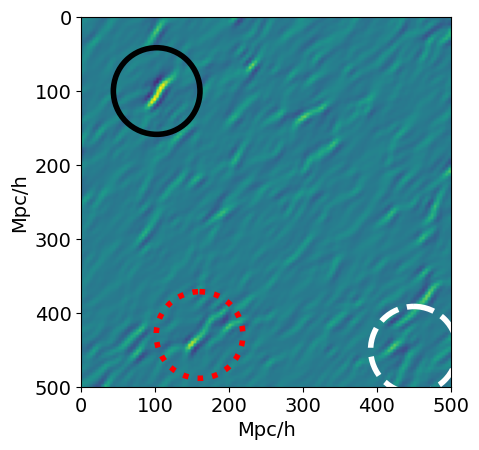} 
  \label{fig:QuijoteFiltered2}
\end{subfigure}
\vspace{-0.55cm}
 \caption{(Left) Typical projected 2D density map of the LSS from the Quijote simulations~\citep{villaescusa2019quijote}. (Center and Right) Real part of the same map convolved with wavelets $\psi_{1,+2}(\vec{x})$ and $\psi_{1,-2}(\vec{x})$, respectively. The dashed white circles highlight filaments captured by the first wavelet, the dotted red circles a filament captured by the second wavelet, and the plain black circles an intersection of filaments captured by both wavelets.}
\label{fig:QuijoteFiltered}
\end{figure*}

\paragraph*{Notation.}
We use $\rho(\vec x)$ to denote the random 2D field under study. We assume that $\rho(\vec{x})$ has homogeneous statistical properties, i.e., that the statistical distribution of the associated process is translation invariant. We also assume that this field has periodic boundary conditions. We work on a Cartesian grid of size $N=256$, so the position $\vec{x}$ is defined in $\left[0,N\right[^2$. The Fourier transform of $A(\vec x)$ is $\hat{A}(\vec{k})$, and $A^*$ is the complex conjugate of $A$. $A*B$ denotes the convolution of $A$ and $B$. The expected value of a stochastic process $X$ is written $\langle X \rangle$, and the covariance between $X$ and $Y$ is $\text{Cov}(X,Y) = \langle X Y^* \rangle - \langle X\rangle\langle Y^*\rangle$.

A public version of the code used in this article is available at \url{https://github.com/Ttantto/wph_quijote}.

\section{Wavelet phase harmonics}
\label{PartWaveletToWPH}

\subsection{Wavelet transform}
\label{PartWavelet}

WPH statistics are based on the wavelet transform, which is an efficient tool for locally separating the multiscale variability of a given process. Wavelets have been used successfully across a wide range of physics research~\citep[see for instance][]{van2004wavelets}. A wavelet transform of a field consists of its convolution with a set of wavelets that probe specific structures. With appropriately chosen wavelets, the transform leads to a sparse spatial description of the structures at different scales. In this paper, we use bump steerable wavelets \cite{mallat2018phase}, which characterize localized directional oscillations and have been used to efficiently synthesize physical fields~\citep{zhang2019maximum}.

The complex bump steerable wavelets $\psi_{j,\ell}(\vec{x})$ are labeled by two integers $j$ and $\ell$. The integer $j$ takes $J$ values from $0$ to $J-1$ and specifies a characteristic wavelength of oscillation of order $2^{j+1}$ in pixel space. In this paper, we take $J=8$ so that this wavelength ranges from 2 to 256 pixels. The integer $\ell$ characterizes the oscillation's orientation, indexing an angle of $2\pi \ell/L$ with respect to the reference axis. In this paper, we divide $2\pi$ into $L = 16$ angles. We can obtain all these wavelets $\psi_{j,\ell}(\vec{x})$ by a dilation and a rotation of one complex mother wavelet $\psi(\vec{x})$:
\begin{equation}
\psi_{j, \ell} (\vec{x}) = 2^{-j} \psi\left(2^{-j} r_{-\ell} \vec{x}\right),
\end{equation}
where $r_{\ell}$ is the rotation of angle $2\pi \ell/L$, and $\psi(\vec{x})$ is defined in Appendix~\ref{AppendixBSWavelets}. Fig.~\ref{FigWavelet2D} shows the real part of such a wavelet, as well as the Fourier transform of one.

The Fourier transform of each bump steerable wavelet $\hat{\psi}_{j,\ell}(\vec{k})$ is real and samples a limited region of the Fourier plane. The mother wavelet is defined with a central frequency $\vec{\xi}_0 = (\xi_0,0)$, and each child wavelet $\psi_{j,\ell}$ has central frequency 
\begin{equation}
\label{EqjellTok}
\vec{\xi} = 2^{-j} r_{\ell} \vec{\xi}_0,
\end{equation}
which we also use as a wavelet index in place of $(j, \ell)$, writing $\psi_{\vec{\xi}}$ instead of $\psi_{j,\ell}$.
When the integers $j$ span all the possible values for a given image (i.e., when $2^J$ is the size of the image), the $\hat{\psi}_{\vec{\xi}}(\vec{k})$ wavelet spectral bands for all $j$ and $\ell$ values cover the whole Fourier plane.

The bump steerable wavelet transform of a field $\rho(\vec x)$ is defined as its convolution with the set of wavelets defined above, that is, the $J\times L$ convolutions $\rho * \psi_{\vec \xi} (\vec x)$. Each of these convolutions corresponds to a local filtering of the field $\rho$ on the frequency support of $\psi_{\vec \xi}$, around the frequency $\vec{\xi}$. Fig.~\ref{fig:QuijoteFiltered} shows two such convolutions on matter density fields of the LSS from the Quijote simulations. Notice how each wavelet picks up the filamentary structures at a given scale and orientation. The values of the resulting filtered fields peak at only a few spatial positions, illustrating the sparsity of the wavelet transform.


\subsection{Covariance of wavelet transforms}
\label{PartCovWavelet}

\begin{figure*}[t]
\begin{center}
\includegraphics[width = 0.99\textwidth]{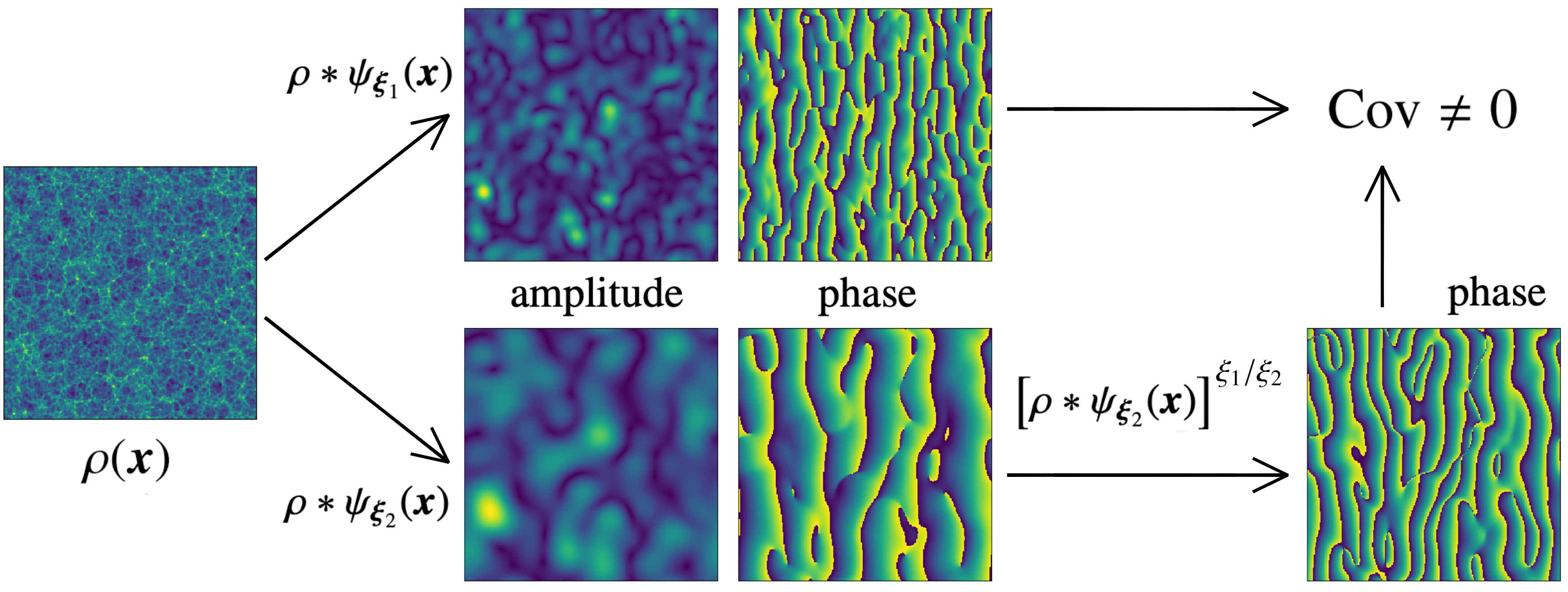}
\vspace{-0.4cm}
\end{center}
 \caption{Illustration of wavelet phase harmonics (WPH) moments computation. A typical Quijote density field $\rho$ (far left) is convolved with two wavelets $\psi_{\vec \xi_1}$ and $\psi_{\vec \xi_2}$, with $(j_1,\ell_1)  = (3,0)$ and $(j_2,\ell_2) = (4,0)$. The amplitude and phase of each convolution is shown in the central panel. From their phase, one sees that the $\rho * \psi_{\vec \xi_i}$ fields oscillate with different characteristic scales $2^{j_1}$ and $2^{j_2}$, respectively. Their covariance is therefore negligible. By applying the phase harmonic operator to $\rho * \psi_{\vec{\xi_2}}$, using harmonic exponent $p = \xi_1/\xi_2 = 2^{j_2}/2^{j_1}$, one obtains a new field of the same amplitude but with a phase of characteristic scale $2^{j_1}$ (lower right). As the fields $\rho * \psi_{\vec \xi_1}$ and $[\rho*\psi_{\vec \xi_2}]^{\xi_1/\xi_2}$ have the same characteristic wavelength, their covariance may be non-negligible. This covariance is a WPH moment characterizing the relative phase alignment between the $\rho*\psi_{\vec \xi_1}$ and $\rho*\psi_{\vec \xi_2}$ fields. This type of WPH moment computation is illustrated in Fourier space in the left panel of Fig.~\ref{FigCouplingFourier}.}
\label{FigConceptWPH}
\end{figure*}

To characterize the dependency between the field $\rho$ filtered at two scales (i.e., $\rho * \psi_{\vec \xi_1} (\vec x)$ and  $\rho * \psi_{\vec \xi_2} (\vec x)$), we could consider the following covariance:
\begin{equation}
\label{EqCovWav}
C_{\vec \xi_1,\vec \xi_2}(\vec{\tau}) =  \text{Cov} \left[\rho * \psi_{\vec \xi_1} (\vec{x}),\rho * \psi_{\vec \xi_2}(\vec{x}+\vec{\tau}) \right].
\end{equation}
For a stationary field this quantity does not depend on $\vec x$ but only on the spatial shift $\vec\tau$.
However, $C_{\vec \xi_1,\vec \xi_2}(\vec{\tau})$ carries no more information than the power spectrum%
\footnote{$S(\vec{k})$ is the complete power spectrum, not the isotropic one. For a stationary process $\rho$, it is defined as the Fourier transform of the two-point correlation function $s(\vec{\tau}) = \text{Cov}\left[\rho(\vec u), \rho(\vec u + \vec \tau) \right]$.}
$S(\vec k)$ of $\rho$ since they are related by~\citep{zhang2019maximum}: 
 \begin{equation}
     C_{\vec \xi_1, \vec \xi_2}(\vec{\tau}) 
     = \int S (\vec{k}) 
     \ \hat \psi_{\vec \xi_1}(\vec{k})
     \ \hat \psi_{\vec \xi_2}^*(\vec{k})
     \ e^{-i \vec{k} \cdot \vec{\tau}} \mathrm{d} \vec{k}.
     \label{eq:linkps2}
 \end{equation}
Eq.~\eqref{eq:linkps2} shows that $C_{\vec \xi_1,\vec \xi_2}(\vec{\tau})$ cannot actually capture the couplings between different scales and angles.  Indeed, this equation implies that $C_{\vec \xi_1,\vec \xi_2}(\vec{\tau})=0$ if the supports of $\psi_{\vec \xi_1}$ and $\psi_{\vec \xi_2}$ in the Fourier plane do not overlap.
 This result is illustrated in Fig.~\ref{FigConceptWPH}, which shows two convolutions of a typical LSS field by wavelets probing different spatial frequencies $\vec \xi_1$ and $\vec \xi_2$. Since the covariance of those maps is basically their scalar product,\footnote{The mean values of the wavelet convolutions vanish, and $\text{Cov}(A,B) = \langle A B^* \rangle$ when $\langle A \rangle = 0 = \langle B \rangle$.} it is negligible because the maps oscillate at different spatial frequencies. 
These results imply that the descriptor $C_{\vec \xi_1,\vec \xi_2}(\vec{\tau})$ cannot capture any coupling between different scales or angles. Thus, this descriptor cannot distinguish between processes that have the same power spectrum, even if their higher order statistics differ. In particular, it cannot distinguish a Gaussian process from a highly non-Gaussian one. 

\subsection{Phase harmonics and coupling between scales}
\label{PartWPH_Concept}

To capture the statistical dependence between non-overlap\-ping scales, we must use nonlinear operators. We therefore use the phase harmonic operator introduced in~\cite{mallat2018phase}:
given a complex number $z$, with modulus $|z|$ and phase $\text{arg}(z)$, its $p$th phase harmonic is defined as
\begin{equation}
\label{EqDefPhaseHarmonic}
\left[z \right]^p = |z| \cdot \text{e}^{i p~  \text{arg}(z)}.
\end{equation}
When applied to a two-dimensional complex map, this operator accelerates the map's spatial frequency of oscillation by a factor $p$, while keeping the modulus unchanged.  This operator therefore modifies the Fourier spectrum of $\rho * \psi_{\vec{\xi}}$ fields without modifying the spatial localization of their features. The spectral band of $\rho * \psi_{\vec{\xi}}$ is localized around frequency $\vec{\xi}$, while the $\left[ \rho * \psi_{\vec{\xi}} \right]^p$ field mainly contains frequencies around $p \vec{\xi}$. Fig.~\ref{FigConceptWPH} illustrates this nonlinear operation: a suitable phase harmonic applied to wavelet transforms $\rho * \psi_{\vec{\xi}_i}$ produces a nonvanishing covariance, which enables the dependency between different scales of the field to be captured.

Therefore we define the \emph{WPH moments} as
\begin{equation}
\label{EqDefWPH_Cov}
\displaystyle{
C_{\vec{\xi}_1,p_1,\vec{\xi}_2,p_2}(\vec{\tau}) = \text{Cov} \left(\left[\rho * \psi_{\vec \xi_1} (\vec{x}) \right]^{p_1},\left[\rho * \psi_{\vec \xi_2} (\vec{x} + \vec{\tau}) \right]^{p_2} \right).
}
\end{equation}
$C_{\vec{\xi}_1,p_1,\vec{\xi}_2,p_2}(\vec{\tau})$ probes the coupling between $\vec{\xi}_1$ and $\vec{\xi}_2$ frequencies. Nonvanishing WPH moments arise only when the frequency bands of $\left[\rho * \psi_{\vec \xi_1} (\vec{x}) \right]^{p_1}$ and $\left[\rho * \psi_{\vec \xi_2} (\vec{x}) \right]^{p_2}$ overlap. This condition is guaranteed if
\begin{equation}
\label{EqSamePHFreq}    
    p_1 \vec{\xi_1} \simeq p_2 \vec{\xi}_2,
\end{equation}
as illustrated in Fig.~\ref{FigConceptWPH}. 
Because of the spectral width of the $\left[ \rho * \psi_{\vec{\xi}} \right]^p$ field, other moments defined by Eq.~\eqref{EqDefWPH_Cov} with overlapping frequency bands can also be constructed when $p_1$ or $p_2$ is equal to zero~\citep{zhang2019maximum}.

A key property of phase harmonics is their \emph{robustness}.  Phase harmonics $[z]^p$ and standard moments $z^p$ capture the same phase couplings. However, the estimation of phase harmonics is more robust than that of standard moments because their modulus is not raised to the $p$th power.\footnote{Indeed, $|[z]^p - [z']^p| \leq \text{max}(|p|,1)|z-z'|$.  See \cite{mallat2018phase}.}
For instance, to couple the characteristic scales of $4$ and $32$ pixels using standard moments, we would have to raise the field to the 8th power, which makes the standard moments extremely susceptible to outliers.
The variance of the WPH moments is bounded more favorably than that of the standard $n$-point statistics (see~\citep{zhang2019maximum} for a theoretical analysis).

The \emph{WPH statistics} introduced in this paper are built from a collection of WPH moments as given in Eq.~\eqref{EqDefWPH_Cov}. Constructing a set of WPH statistics boils down to selecting an ensemble of WPH moments, which are specified by $(\vec \xi_1, p_1,\vec \xi_2,p_2)$ parameters. We tailor this selection depending on the field $\rho$ and the purpose of the statistics.

\subsection{Symmetries and spatial shift discretization}
\label{PartSymmetries}

\paragraph*{Symmetries and invariant WPH description.}

If the physical phenomenon under study possesses some symmetries (i.e., if its statistical properties are invariant under certain groups of transformations), we can take them into account and lower the dimension of the WPH statistics. Note that we have already implicitly assumed that the field is invariant under translation when we defined $C_{\vec{\xi}_1,p_1,\vec{\xi}_2,p_2}(\vec{\tau})$.

If the field is invariant under rotations, the WPH moments have angular dependency on only $\delta_\ell = \ell_2-\ell_1$. Similarly, the field may be invariant under parity, which corresponds in two dimensions to invariance when one of the axes of an image is flipped. This symmetry also expresses that a clockwise and an anticlockwise rotation cannot be distinguished. When both parity and rotational invariance hold, the WPH moments depend on only $|\delta \ell| = |\ell_2-\ell_1|$.

The matter density field from Quijote simulations is expected to be invariant under translations, rotations, and parity. These symmetries allow construction of parity-invariant WPH moments, which we label $\mathcal{C}^{\text{isopar}}$ and define as 
\begin{equation}
\label{EqPhaseCoupling_Iso}
\displaystyle{
{\mathcal{C}}^{\text{isopar}}_{j_1,p_1,j_2,p_2,\delta \ell}(\vec{\tau}) = 
\left \langle 
\mathcal{C}_{j_1,\ell_1,p_1,j_2,\ell_2, p_2} (\vec{\tau})
\right \rangle_{  |\ell_2-\ell_1| = \delta\ell} },
\end{equation}
where $\langle \; \rangle $ stands for an angular average (over $\ell_1$ and $\ell_2$), $\delta \ell \ge 0$, and the moment $\mathcal{C}_{j_1,\ell_1,p_1,j_2,\ell_2, p_2}$ refers to the standard WPH moment $\mathcal{C}_{\vec\xi_1,p_1,\vec\xi_2,p_2}$ of Eq.~\eqref{EqDefWPH_Cov} with the $\vec\xi$ and $(j,\ell)$ indices related by Eq.~\eqref{EqjellTok} in the usual way. These invariant $\mathcal{C}^{\text{isopar}}$ moments significantly reduce the dimension of the WPH statistics, which reduces the variance of their estimators.

\paragraph*{Discretization of spatial shift and spectral resolution.}

Since a convolved field $\rho * \psi_{\vec \xi_i}$ has been filtered at a $2^{j_i+1}$ scale, we gain little or no additional information from sampling it at a finer scale. This allows us to consider discrete sets of translations $\vec{\tau}$, which also is in accord with the discretized approach in general. We use different sets of translations depending on the application (see Appendix~\ref{AppendixFinalModels}). Using a large number of spatial shifts $\vec{\tau}$ improves the spectral resolution of the statistics, but increases the number of WPH moments. There is therefore a trade-off between the number of WPH moments and the spectral resolution; see Sec.~\ref{PartAnalysisL1L2} for further discussion.

\smallbreak
\paragraph*{WPH statistics used in this paper.}
\label{SecIntermediateConclusion}

%
Defining a set of WPH statistics for a particular field and purpose amounts to selecting a set of $\{\vec \xi_1, p_1, \vec \xi_2, p_2, \vec \tau\}$ parameters and a set of symmetries (such as rotational invariance).
%
Appendix~\ref{AppendixFinalModels} describes the specific WPH statistics used in this paper, first to estimate Fisher information about cosmological parameters in Sec.~\ref{PartCosmoInfo} and then to produce realistic statistical syntheses in Sec.~\ref{PartSyntheses}.
We base our choice of WPH moments on numerical experiments and also on the physical interpretation of the WPH moments, as will be discussed in Sec.~\ref{PartDiscussionVariousTerms}.

\section{Fisher information on cosmological parameters}
\label{PartCosmoInfo}


We evaluate in this section the ability of WPH statistics to infer cosmological parameters. To do so, we compute their Fisher information with respect to five cosmological parameters for 2D matter density fields from the Quijote simulations. First, we describe the Quijote simulations and the density fields used. Next, we outline the Fisher analysis that is performed. Finally, we show how our results using WPH statistics compare to state-of-the-art results obtained with two widely used summary statistics: the standard power spectrum, and the joint power spectrum and bispectrum. 

\subsection{Quijote simulations}
\label{PartQuijoteSimulations}

\begin{table}[t]                        
\centering
\begin{tabular}{| c | c | c | c | c | c | }
\hline
\xrowht{17.5pt}
Parameter & $\Omega_m$ & $\Omega_b$ & $h$ & $n_s$ & $\sigma_8$  \\
\hline
\xrowht{17.5pt}
$\theta_\alpha^\text{fid}$ & 0.3175 & 0.049 & 0.6711 & 0.9624 & 0.834 \\
\hline
\xrowht{17.5pt}
$\Delta \theta_\alpha$ & 0.01 & 0.002 & 0.02 & 0.02 & 0.015 \\
\hline
\end{tabular}
\caption{Fiducial values $\theta_\alpha^\text{fid}$ and finite deviations $\Delta \theta_\alpha$ of the cosmological parameters used in our simulations.}
\label{TableCosmoParam}    
\end{table}

The physical LSS field we study in this paper is the spatial distribution of the underlying matter density field, $\rho(\vec{x})$, which we obtain from the Quijote simulations~\citep{villaescusa2019quijote}. The Quijote simulations are a set of 43100 full N-body simulations of the LSS, tracing the evolution of spatial fluctuations from redshift $z=127$ to $z=0$. The initial conditions at $z=127$ are computed using 2LPT with CAMB~\citep{Lewis:1999bs}, while the dynamics of the simulations that follows the evolution of the dark matter particles relies on the TreePM+SPH code Gadget-III, an improved version of Gadget-II~\citep{Springel:2005mi}. See \cite{villaescusa2019quijote} for further details on these simulations.

In this paper, we use 2D matter fields of 256$\times$256 pixels, which are generated as follows: First, for each realization we compute a 3D density field with $256^3$ voxels by assigning particle positions to the grid using the cloud-in-cell mass assignment scheme. Then we take a slice of $256\times256\times64$ and project to $256^2$ pixels by computing the average along the third axis. The resulting field represents a region with an area of 1000$\times$1000 $(h^{-1}{\rm Mpc})^2$. The matter density fields $\rho(\vec{x})$ are normalized to satisfy $\overline{\rho} = 1$. In the following section, we study both the matter density field and its logarithm. Figure~\ref{FigConceptWPH} provides an example of such a field.

We consider different cosmologies with five varying cosmological parameters: the matter density parameter $\Omega_m$, the baryon density parameter $\Omega_b$, the dimensionless Hubble parameter $h$, the scalar spectral index $n_s$, and finally $\sigma_8$, the average rms matter fluctuation smoothed at $8h^{-1}$Mpc scale. We denote these parameters collectively as $\theta_\alpha$.

We use two different sets of simulations. The first set contains 15000 simulations of the Planck fiducial cosmology~\citep{aghanim2018planck}, for cosmological parameters $\theta_\alpha^\text{fid}$ (see Table~\ref{TableCosmoParam}). The second set of simulations is devised to numerically compute partial derivatives with respect to the five cosmological parameters. For each cosmological parameter $\theta_\alpha$, this set contains 1000 simulations for $\theta_\alpha^\text{fid} \pm \Delta \theta_\alpha$ (i.e., 500 for each sign), the other parameters being held fixed at the fiducial values (see \cite{villaescusa2019quijote} for more details). See Table~\ref{TableCosmoParam} for the values of $\Delta \theta_\alpha$. 

\begin{table}[t]                       
\centering
\begin{tabular}{| c | c | c | c | c | c | c |}
\hline
\xrowht{17.5pt}
$\Phi$ & {\small P$_k$}  &  {\small P$_k$ + B$_k$}  & {\small WPH}
& {\small P$_k^\prime$} &  {\small P$_k^\prime$ + B$_k^\prime$} & {\small WPH$^\prime$} \\
\hline
\xrowht{17.5pt}
Size & 127 & 313 & 327 & 127 & 313 & 327 \\
\hline
\hline
\xrowht{17.5pt}
$\Omega_m$ & 0.15  & 0.12 & 0.11
& 0.15  & 0.12 & 0.10 \\
\hline
\xrowht{17.5pt}
$\Omega_b$ & 0.16 & 0.12 & 0.075
& 0.12  & 0.097 & 0.064  \\
\hline
\xrowht{17.5pt}
$h$ & 1.5 & 1.1 & 0.71
& 0.99 & 0.78 & 0.50  \\
\hline
\xrowht{17.5pt}
$n_s$ & 0.74  & 0.52 & 0.20
& 0.25 & 0.20 & 0.11  \\
\hline
\xrowht{17.5pt}
$\sigma_8$ & 0.024 & 0.013 & 0.018
& 0.012 & 0.0097 & 0.0097  \\
\hline
\end{tabular}
\caption{Marginalized errors on cosmological parameters obtained with Fisher analysis of the matter density field (columns 2 to 4) and its logarithm (primes; final three columns). The analysis used power spectrum (P$_k$), joint power spectrum and bispectrum (P$_k$ + B$_k$), and WPH statistics.}
\label{TableFisherResultsMain}     
\end{table}

\subsection{Fisher matrix analysis}

For a given set of statistics, we can quantify the information that they contain (on average) with respect to the cosmological parameters $\theta_\alpha$ by computing the Fisher information matrix of the parameters.
Specifically, consider a set $\{\Phi_1(\rho), \ldots, \Phi_d(\rho)\}$  of $d$ scalar statistics (such as WPH moments) computed from a realization $\rho$ of the field.
We denote by $\mu_i(\theta_\alpha)$ the expected value of $\Phi_i(\rho)$, and by $\Sigma_{ij}(\theta_\alpha)$ the covariance of $\Phi_i(\rho)$ and $\Phi_j(\rho)$
when $\rho$ is drawn under $\theta_\alpha$.
If the statistics are jointly Gaussian and their covariance matrix $\Sigma$
does not
depend on $\theta_\alpha$, the Fisher information matrix  boils down to~\citep{tegmark1997karhunen}
\begin{equation}
    \label{EqFisherMatrix}
    F_{\alpha\beta} 
    = 
    \sum_i\sum_j
    \frac{\partial \mu_i}{\partial \theta_\alpha}  \ 
    \left(\Sigma^{-1}\right)_{ij} \
    \frac{\partial \mu_j}{\partial \theta_\beta} .
\end{equation} 
From this Fisher matrix we can compute the Cram\'er-Rao bound, which gives the asymptotically lowest possible variance $\delta \theta_\alpha^2$ for any unbiased estimator of $\theta_\alpha$ based on $\Phi$:
\begin{equation}
    \label{EqCramerRao}
    \delta \theta_\alpha \geq \sqrt{{\bigl(F^{-1}\bigr)_{\alpha\alpha}}}.
\end{equation}
In this paper, we numerically estimate the Fisher matrices (for each set of summary statistics considered) for the cosmological parameters $\theta_\alpha$ corresponding to the Planck fiducial cosmology. We estimate the covariance matrices from the 15000 fiducial Planck simulations, while each partial derivative appearing in Eq.~\eqref{EqFisherMatrix} is evaluated with the two sets of 500 simulations at $\theta_\alpha^\text{fid} \pm \Delta \theta_\alpha$. We checked the convergence of these estimates by verifying that the Cram\'er-Rao bounds changed only at the percent level when using 10000 (350) simulations to compute the covariance matrices (partial derivatives).

\begin{table}[t]               
\centering
\begin{tabular}{| c | c | c | c | c | c | c | }
\hline
\xrowht{17.5pt}
$\Phi$ &\small{WPH}&\small{+ P$_k$}&\small{+ B$_k$}&\small{WPH$^\prime$}&\small{+ P$_k^\prime$}&\small{+ B$_k^\prime$}\\
\hline
\xrowht{17.5pt}
Size &  327 & 454 & 513
& 327 & 454 & 513  \\
\hline
\hline
\xrowht{17.5pt}
$\Omega_m$ & 0.11 & 0.11 & 0.10
& 0.102 & 0.096  & 0.094  \\
\hline
\xrowht{17.5pt}
$\Omega_b$ & 0.075 & 0.073 & 0.070 
& 0.064 & 0.063  & 0.062  \\
\hline
\xrowht{17.5pt}
$h$ & 0.71 & 0.68 & 0.65
& 0.50 & 0.50  & 0.48 \\
\hline
\xrowht{17.5pt}
$n_s$ & 0.20 & 0.20 & 0.19
& 0.11 & 0.11  & 0.11  \\
\hline
\xrowht{17.5pt}
$\sigma_8$ & 0.018 & 0.018 & 0.0095 
& 0.0097 & 0.0096  & 0.0086  \\
\hline
\end{tabular}
\caption{Marginalized errors on cosmological parameters obtained with Fisher analysis using WPH, joint WPH + P$_k$, and joint WPH + B$_k$ statistics. These results were obtained with the matter density field (columns 2 to 4), and its logarithm (primes; final 3 columns).}
\label{TableFisherResults2}             
\end{table}

\begin{figure*}[t]
\begin{center}
\includegraphics[width = 0.95\textwidth]{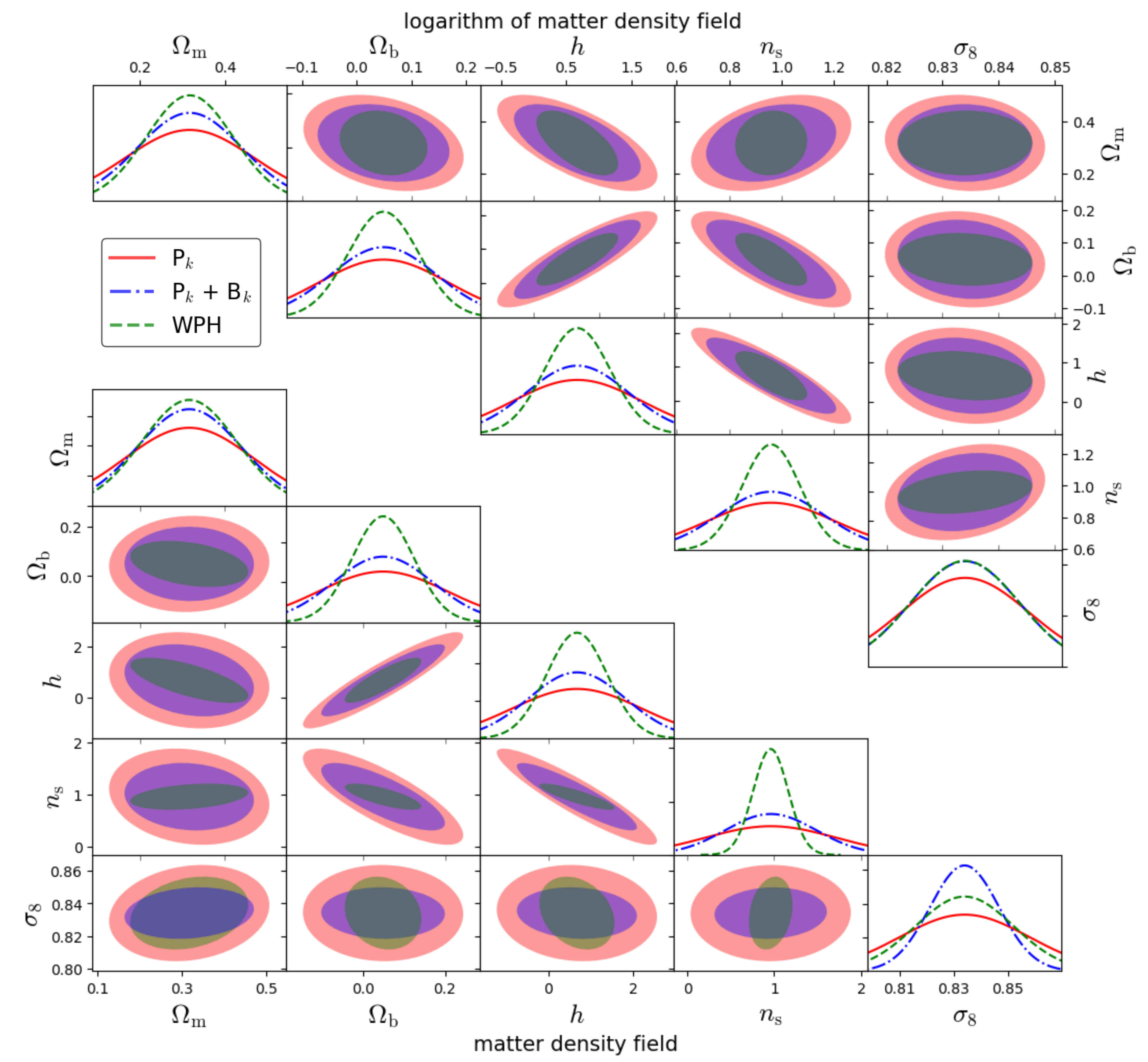}
\vspace{-0.4cm}
\end{center}
\caption{Fisher matrix constraints for five cosmological parameters, based on WPH statistics (green), power spectrum (P$_k$, red), and joint power spectrum and bispectrum (P$_k+$B$_k$, blue). These constraints were computed from (1 Gpc/h)$^2$ maps of the projected matter density field (bottom) and its logarithm (top). Contours mark 95$\%$ confidence intervals. WPH statistics provide the best constraints for each parameter except $\sigma_8$, which is more tightly constrained by P$_k+$B$_k$ on the matter density field (bottom row).}
\label{FigFisher}
\end{figure*}

\subsection{Fisher matrix results}

We compare in this section the results obtained with three sets of summary statistics: the standard isotropic power spectrum (P$_k$), the power spectrum plus a set of isotropic bispectrum triangles (P$_k$ + B$_k$), and a set of WPH statistics. 
Appendix~\ref{AppendixBSTriangle} describes the bispectrum triangle ensemble, which contains flattened, squeezed, and equilateral triangles.
The WPH statistics are constructed from the WPH moments given in Eq.~\eqref{EqPhaseCoupling_Iso}, which are invariant under rotations and parity. Appendix~\ref{AppFisherWPHModel} provides a complete description of these moments, which characterize all the scales of the image with 327 coefficients. A plot of a subset of moments from these WPH statistics are given in Fig.~\ref{FigPlotCoeffs}, and their covariance in Fig.~\ref{FigPlotCorr}. Note however that these figures use notations introduced in Sec.~\ref{PartDiscussionVariousTerms}.

For each of these descriptors, we evaluated the posterior distributions of the cosmological parameters obtained using the matter density field and its logarithm from the Quijote simulations. We expect the logarithm to make the density field more Gaussian, while transferring information from high-order correlations to the power spectrum~\citep{neyrinck2009rejuvenating,neyrinck2011rejuvenating,neyrinck2011rejuvenatingIII}. See~\cite{massara2020using} for a similar case of a nonlinear transform applied to Quijote simulations.

Table~\ref{TableFisherResultsMain} and Fig.~\ref{FigFisher} respectively show the marginal and full posterior distributions of cosmological parameters obtained with the Fìsher analysis. For all five cosmological parameters, the WPH statistics contain more information than the power spectrum. The improvement of forecast errors ranges from $20\%$ to a factor larger than 3. The relative improvement is generally larger for the matter density field than for its logarithm. This may be because the logarithm of the field is more Gaussian and the power spectrum suffices to characterize a Gaussian stationary field.

Compared to the power spectrum plus bispectrum, the WPH statistics provide better constraints on four of the cosmological parameters, the exception being $\sigma_8$. For those four parameters, the absolute improvement from P$_k$ + B$_k$ to WPH is similar to the improvement from P$_k$ alone to P$_k$ + B$_k$.

Table~\ref{TableFisherResults2} shows the Fisher information results for joint WPH + P$_k$ and WPH + B$_k$ statistics. Except for $\sigma_8$ with WPH + B$_k$ statistics, only limited additional information is gained by adding these statistics to those built solely on WPH moments. 

\begin{figure*}[t]
\begin{center}
\includegraphics[width = 0.99\textwidth]{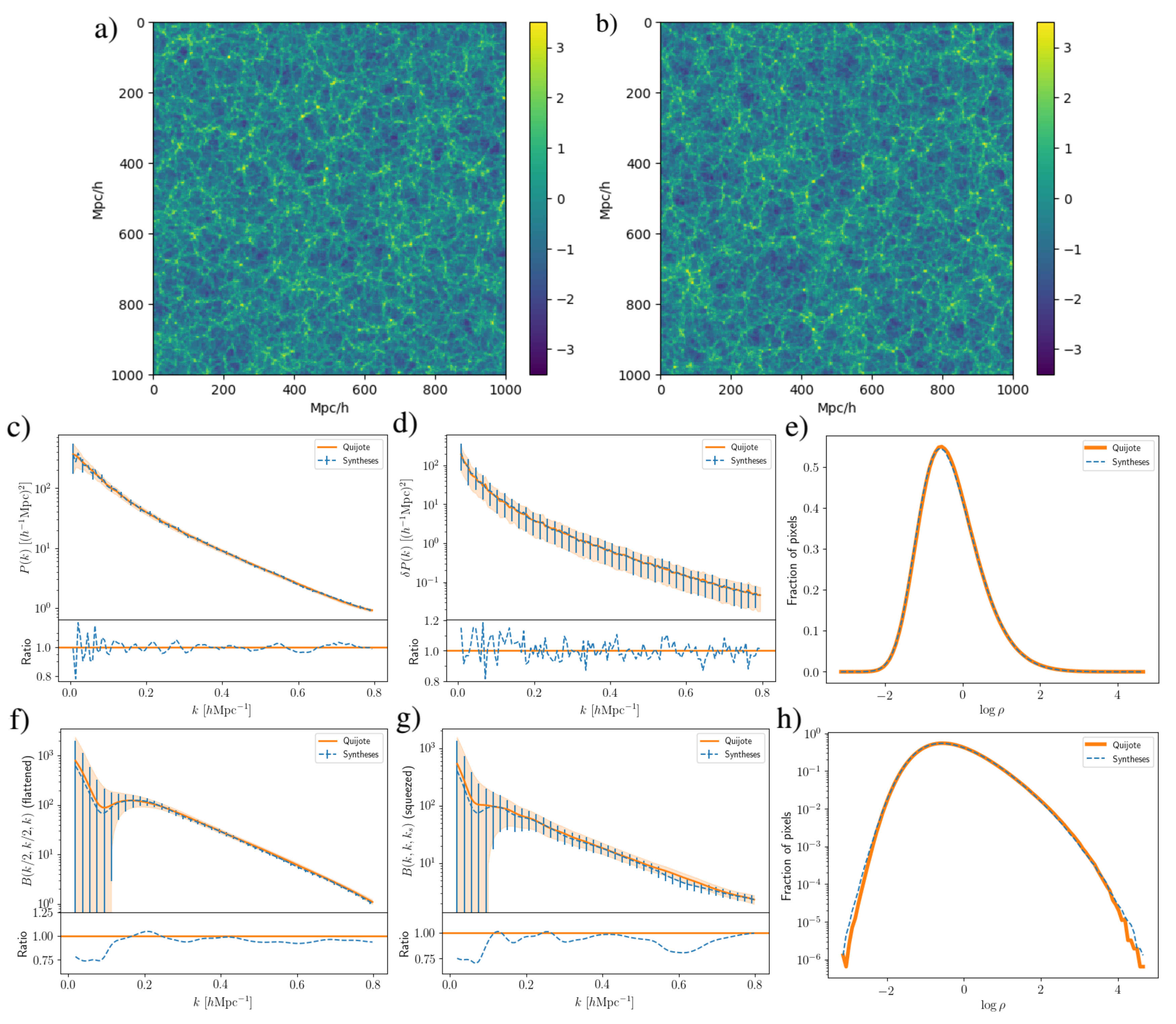}
\vspace{-0.6cm}
\end{center}
 \caption{
 Comparisons of the logarithm of the matter density field $\log(\rho)$ in Quijote simulation maps and in our statistically synthesized maps, showing how well the syntheses reproduce the statistical properties of $\log(\rho)$ in the Quijote maps. The error bars correspond to the realization-per-realization dispersion.
 a) A map of $\log(\rho)$ from the Quijote simulations. b) A map of $\log(\rho)$ synthesized based on WPH statistics of a sample of 30 Quijote maps (see Sec.~\ref{sec:Algorithm}). 
 c)--h) Statistics for $\log(\rho)$ estimated using 300 maps from the Quijote simulations (orange lines) and 300 syntheses (dashed blue lines).
 c) Power spectrum, 
 d) standard deviation of the power spectrum, 
 e) pixel value PDF on a linear scale, 
 f) bispectrum in the flattened triangle configuration, $B(k/2, k/2, k)$,
 g) bispectrum in the squeezed triangle configuration, $B(k, k, k_3)$, for $k_3 \ll 1$, and 
 h) pixel value PDF on a logarithmic scale. 
 }
\label{FigResultSyntheses1}
\end{figure*}

Numerous works~\citep[e.g.,][]{Byun:2017fkz} study the LSS field using the power spectrum together with bispectrum. For instance,~\cite{sefusatti2006cosmology,yankelevich2019cosmological,chudaykin2019measuring} provide bispectrum forecasts for full sets of cosmological parameters. In particular,~\cite{hahn2019constraining} computes from the N-body Quijote simulations the full information content of the redshift-space halo bispectrum for six cosmological parameters. However, the particular two-dimensional projected matter density field that we use here makes it difficult to quantitatively compare our results with those obtained in these earlier works. Reference \cite{coulton2019constraining} performs a similar analysis for weak-lensing surveys. They show the improvements gained by adding the bispectrum to the power spectrum and obtain results similar to our results for P$_k$ and P$_k$ + B$_k$.
Given the widespread use of the bispectrum in cosmological parameter inference, current bispectrum results can be taken as a generic benchmark. Thus, we claim that the results obtained with WPH statistics compare favorably to state-of-the-art results obtained with these other summary statistics. 
A  more quantitative comparison is deferred to later work, since it requires WPH statistics to be extended to 3D fields.

\section{Statistical syntheses with WPH statistics}
\label{PartSyntheses}


In this section, we show that WPH statistics embed a wide range of summary statistics commonly used in cosmology. For that purpose, we estimate WPH statistics on a subset of Quijote simulation maps, and generate from them synthetic maps based on a microcanonical maximum entropy principle. Then we compare these statistical syntheses to the whole sample of Quijote maps, using as summary statistics the power spectrum, bispectrum, probability density function (PDF), and Minkowski functionals.

Previous studies have considered similar syntheses based on either WPH statistics~\citep{zhang2019maximum, villaescusa2019quijote} or the related Wavelet Scattering Transform \citep{mallatWST, BrunaMallatWST,allys2019rwst}, but they reported only qualitative assessments, mostly from visual inspection.  In contrast, in Sect.~\ref{sec:resultSynth} below we report much more stringent quantitative tests based on a wide set of statistics commonly used in astrophysics.

\subsection{Microcanonical maximum entropy model}
\label{sec:Algorithm}

In this section we outline our generative algorithm for drawing sample realizations of a microcanonical maximum entropy model. 
A maximum entropy model is a probability distribution $p$ that satisfies a set of statistical constraints, while being \emph{as general as possible} otherwise. This means that it maximizes the Shannon entropy $H(p) = -\int p(\rho) \log\left[ p(\rho)\right] ~ \dd \rho$.
In this paper, we consider microcanonical models, which are defined as follows.
Let $\tilde{\rho}$ denote a realization of the process under study and let $\Phi(\tilde{\rho})$ be a set of statistics computed on this realization.
We define the microcanonical set $\Omega_\varepsilon$ of width $\varepsilon$ conditioned by $\tilde{\rho}$ as
\begin{equation}
    \label{EqMicroSet}
    \Omega_\varepsilon 
    = \bigl\{
    \rho: d\left[\Phi(\rho),\Phi(\tilde \rho)\right] \leq \varepsilon \bigr\},
\end{equation}
where $d\left[\Phi(\rho),\Phi(\tilde \rho)\right]$ is a measure of the discrepancy between $\Phi(\rho)$ and $\Phi(\tilde \rho)$.  The microcanonical maximum entropy model is the model of maximal entropy defined over $\Omega_\varepsilon$. This implies that it has a probability distribution that is uniform on $\Omega_\varepsilon$. See~\cite{bruna2018multiscale} and~\cite{zhang2019maximum} for the exact definition of $d\left[\Phi(\rho),\Phi(\tilde \rho)\right]$ and further explanation.

We could sample from microcanonical maximum entropy models with Monte-Carlo techniques, but such methods tend to be quite computationally expensive for a large number of statistical constraints~\citep{lustig1998microcanonical}. 
We therefore rely on a different approach, introduced in~\citep{bruna2018multiscale}.
To produce one realization of the microcanonical model, we start from a realization $\rho$ of an homogeneous and isotropic Gaussian field which is then modified iteratively by gradient descent with respect to the loss function $\mathcal{L} = d\left[\Phi(\rho),\Phi(\tilde \rho)\right]$. Care must be taken that the descent preserves the key symmetries of the starting point: homogeneity and isotropy.

For this paper, we implement the microcanonical maximum entropy sampling in Python using the PyTorch library~\cite{paszke2017automatic} to compute the gradient of the loss and we perform the loss descent using the L-BFGS-B~\cite{optimizer} implementation of Scipy~\cite{2020SciPy-NMeth}.

\begin{figure*}[t]
\begin{subfigure}{.48\textwidth}
\begin{center}
\includegraphics[width = 0.99\textwidth]{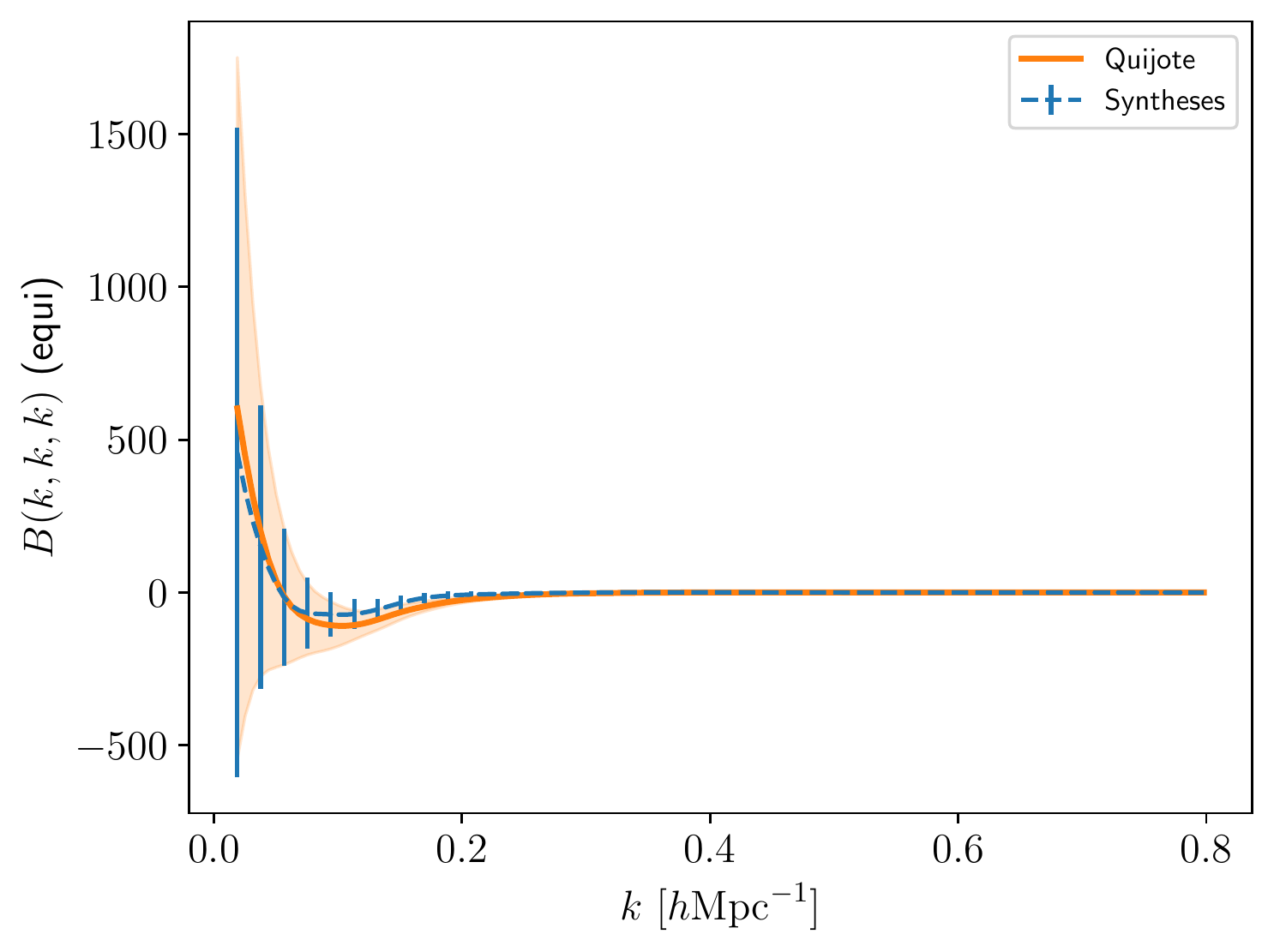}
\vspace{-0.15cm}
\end{center}
\end{subfigure}
\begin{subfigure}{.48\textwidth}
\begin{center}
\includegraphics[width = 0.99\textwidth]{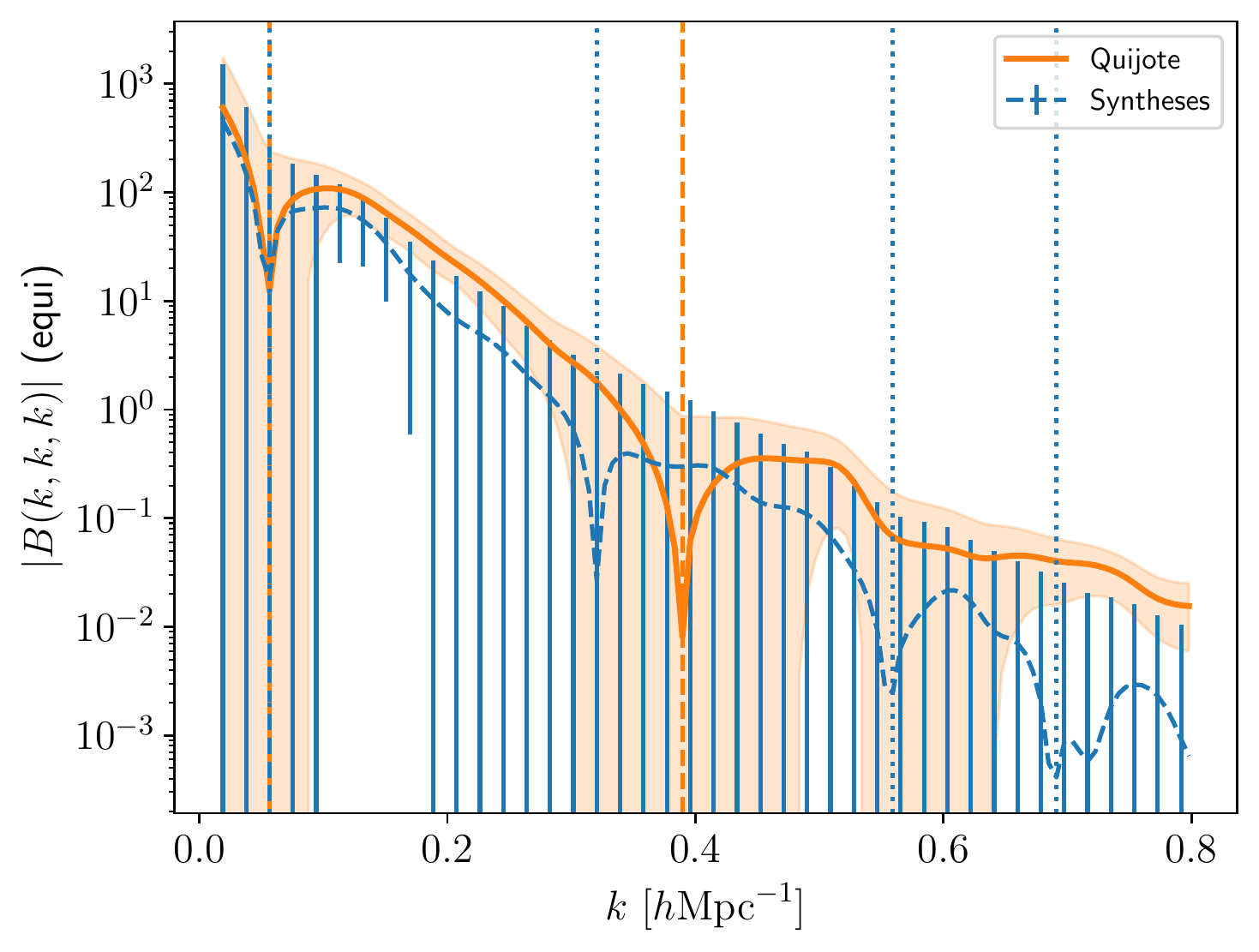}
\vspace{-0.15cm}
\end{center}
\end{subfigure}
\caption{Left: Comparison of the bispectrum of $\log(\rho)$ in the equilateral configuration computed from 300 Quijote simulations (orange) and 300 syntheses (dashed blue). The error bars correspond to the realization-per-realization dispersion. Right: plot of the absolute value of the same quantities in logarithmic scale. Orange dashed vertical lines show the change of sign of the bispectrum of the Quijote simulation and blue dotted lines the change of sign of the bispectrum of the syntheses. Please notice the change of the vertical order of the lines due to the absolute value.}
\label{FigBkEqui}
\end{figure*}

\subsection{Statistical validation of the syntheses}
\label{sec:resultSynth}

We now assess the quality of the syntheses generated by the maximal entropy model. Rather
than working with the matter density field itself, we chose to work with its logarithm.
As the matter density field roughly follows a log-normal distribution, the distribution of
its logarithm should be well approximated by a normal distribution.  Indeed, we found that
it was better reproduced by our maximal entropy model. We refer to
Appendix~\ref{AppLimitations} for further explanations on this choice.

The WPH statistics used in this section, and detailed in
Appendix~\ref{AppSynthesesWPHModel}, contain 6676 coefficients, significantly more than
the 327 coefficients used for Fisher analysis in Sec.~\ref{PartCosmoInfo}.
The increase occurs because the WPH moments used here are not invariant under rotations:
they are constructed from Eq.~\eqref{EqDefWPH_Cov} rather than from the isotropized
version~\eqref{EqPhaseCoupling_Iso}.
We found that this choice led to better syntheses: even though syntheses based on
isotropic moments~\eqref{EqPhaseCoupling_Iso} are visually indistinguishable from syntheses based on anisotropic moments~\eqref{EqDefWPH_Cov}, the various validation statistics presented below are better reproduced with anisotropic moments.
This is likely due to the Cartesian grid breaking the rotational symmetry, especially at
small scales.  The other difference from the statistics of Sec.~\ref{PartCosmoInfo} is
that here we use low-pass filters rather than WPH moments to constrain the scales at
$j\ge 6$.
 
\def\Nlearn{{N_{\textrm{learn}}}}
\def\Nbatch{{N_{\textrm{batch}}}}
 
As sketched in section~\ref{sec:Algorithm}, the principle of the synthesis of a map $\rho$
is to adjust its pixels in order for its WPH moments $\Phi(\rho)$ to match those estimated
from Quijote simulations.  However, something slightly different is implemented in
practice so some details are in order.
Regarding the target WPH moments, they are collected in a vector $\Phi_\textrm{target}$ obtained by averaging (a sample version of) Eq.~\eqref{EqDefWPH_Cov} over a set of $\Nlearn=30$ Quijote maps with periodic boundary conditions. Each Quijote map has a surface area of 1~(Gpc/$h$)$^2$ and is sampled on a grid of $256\times 256$ pixels.  
We found empirically that a set of $\Nlearn=30$ maps was large enough to estimate the  WPH coefficients up to $J=6$ with an accuracy sufficient for our purposes.
We could have used a larger training set but we restrained ourselves to $\Nlearn = 30$ in order to illustrate that our method performs well with a small number of examples.

Regarding the synthesis process itself, maps are not produced individually but in batches
of $\Nbatch$ maps.  We start from $\Nbatch$ maps $\rho_1,\ldots,\rho_\Nbatch$ of size
$256\times256$ obtained as independent Gaussian white noise realizations.  Then, their
pixel values are adjusted by minimizing the joint loss
\begin{equation}\label{eq:jointloss}
  \mathcal{L}(\rho_1,\ldots,\rho_\Nbatch\bigr)
  =\sum_{i=1}^{\Nbatch} 
  d\Bigl(\Phi(\rho_i),\, \Phi_\mathrm{target}\Bigr)
\end{equation}
rather than minimizing the individual loss $d\bigl(\Phi(\rho_i), \Phi_\mathrm{target}\bigr)$ independently for each map $\rho_i$.
The motivation for this variant is that criterion~(\ref{eq:jointloss})
demands that the WPH statistics be matched \emph{on average} over
$\Nbatch$ maps rather than for every map, thereby allowing some
variability in the synthesis\footnote{Such a variability is needed at large
scales which have much fewer degrees of freedom. This issue will be
the subject of future research}.
In this paper, we chose $\Nbatch=30$ because it worked well in our
syntheses and also because it was the practical upper limit imposed
by the size of GPU memory.
For our experiments, we ran 10 optimizations of~(\ref{eq:jointloss})
using the same $\Phi_\mathrm{target}$, hence producing $10\times \Nbatch=300$
synthetic maps, taking about 50 GPU hours\footnote{The GPU used was a
  GPU Nvidia Tesla P100 with 16Go of RAM.}.
Fig.~\ref{FigResultSyntheses1}-a,b shows one Quijote map and one synthesized map.

\begin{figure*}[t]
\begin{center}
\includegraphics[width = 0.99\textwidth]{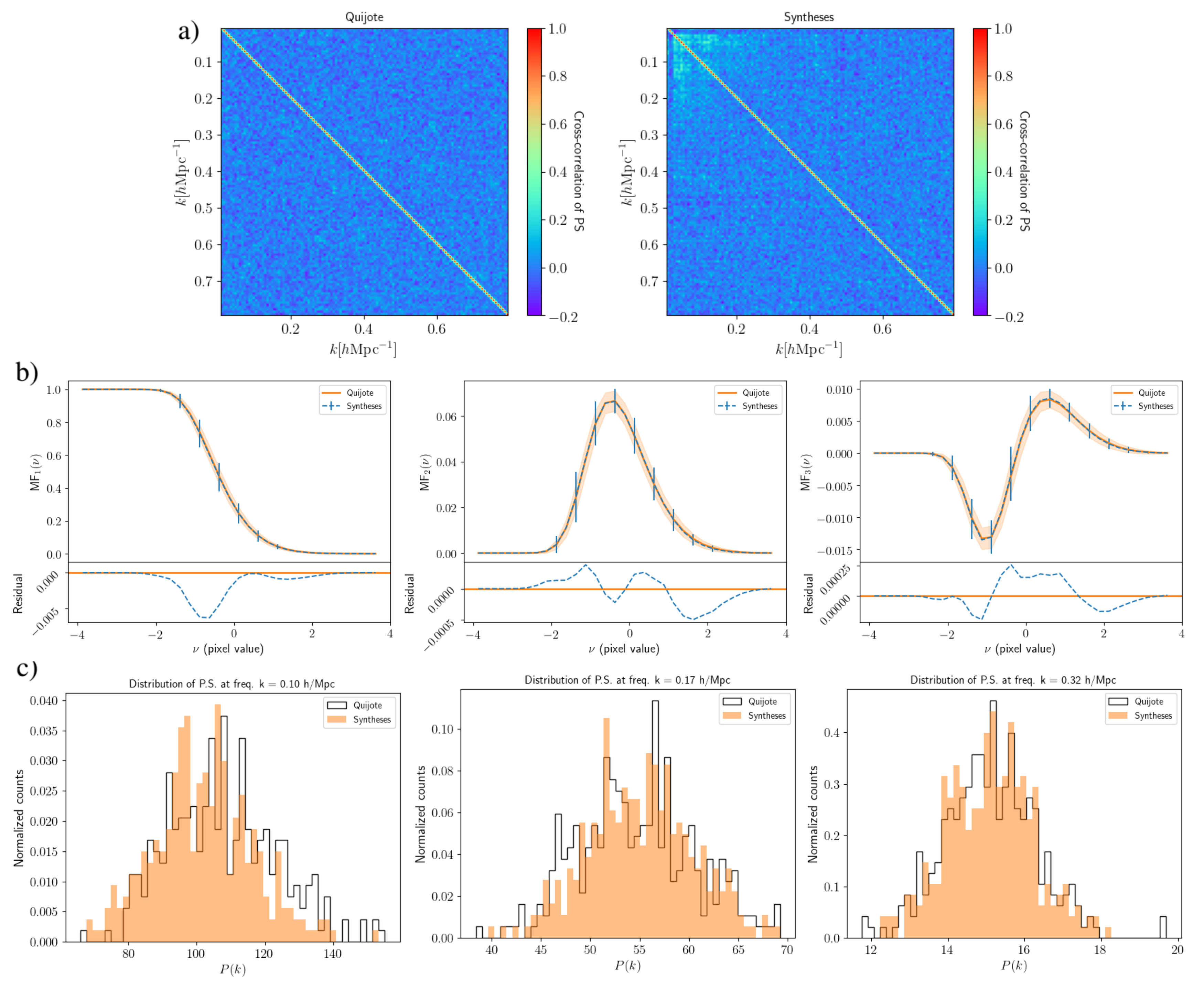}
\vspace{-0.6cm}
\end{center}
 \caption{Further comparisons of $\log(\rho)$ in 300 Quijote maps and our 300 syntheses, showing how well the syntheses reproduce the statistical properties of the Quijote maps.
 a) Correlation matrix of the power spectrum of $\log(\rho)$ in Quijote maps (left) and syntheses (right). Notice that this matrix is mainly diagonal because we consider $\log(\rho)$.  The correlation matrix of the power spectrum of $\rho$ is highly non-diagonal, as also seen on Fig~(3) of~\cite{villaescusa2019quijote}.
 b) First, second, and third Minkowski functionals of Quijote maps (thick orange lines) and syntheses (dashed blue lines). The error bars correspond to the realization-per-realization dispersion. The error bars have been inflated by a factor 5 in these three plots in order to be visible.
 c) Histograms of the distribution of the power spectrum, at the frequencies $k= 0.1$, $0.17$, and $0.32$~h/Mpc for Quijote maps (orange) and syntheses (blue). 
 }
\label{FigResultSyntheses2}
\end{figure*}

\medskip

To assess the quality of the syntheses, we performed a variety of statistical analyses, comparing the 300 synthetic maps to an independent sample of 300 Quijote maps. In particular, we computed several $n$-point statistics for $n$ up to 4 on both these sets of maps.
We compared the pixel distributions ($n=1$), power spectra ($n=2$), bispectra ($n=3$), and the standard deviation and correlation matrix of the empirical power spectrum ($n=4$).
We also computed three Minkowski functionals (MFs) for both sets of 300 maps.
Figs.~\ref{FigResultSyntheses1}, \ref{FigBkEqui}, and~\ref{FigResultSyntheses2} present these results.

For isotropic homogeneous fields, the bispectrum is defined by three wave-vectors $(\vec k_1, \vec k_2, \vec k_3)$ satisfying the triangle inequalities, and therefore representing the lengths of the edges of a triangle. We focus on three configurations: squeezed triangles ($\vec k_1 \simeq \vec k_2$ and $k_3 \ll k_1$), flattened triangles ($\vec k_1 = \vec k_2 = \vec k_3/2$), and equilateral triangles ($k_1 = k_2 = k_3$). 
Fig.~\ref{FigResultSyntheses1}-f,g shows the results for flattened and squeezed triangles and Fig.~\ref{FigBkEqui} those for equilateral triangles. Appendix~\ref{AppendixBSTriangle} describes the bispectrum computation in greater detail.

Minkowski functionals are statistics capturing the topology of the level sets of the field.  They are used in cosmology to probe the non-Gaussianity of the CMB \citep{Ade:2013nlj}, to probe departures from General Relativity of the LSS \citep{Fang:2017daj}, and to study lensing convergence maps \citep{Parroni:2019snd}. In $2$ dimensions, there are three MFs that depend on a threshold $\nu$, denoted $V_0(\nu)$, $V_1(\nu)$ and $V_2(\nu)$. Their definitions are recalled in appendix \ref{AppMF}. The MFs of the sets of maps are shown in figure~\ref{FigResultSyntheses2}-b.

\begin{figure*}[t]
\begin{center}
\includegraphics[width = 0.99\textwidth]{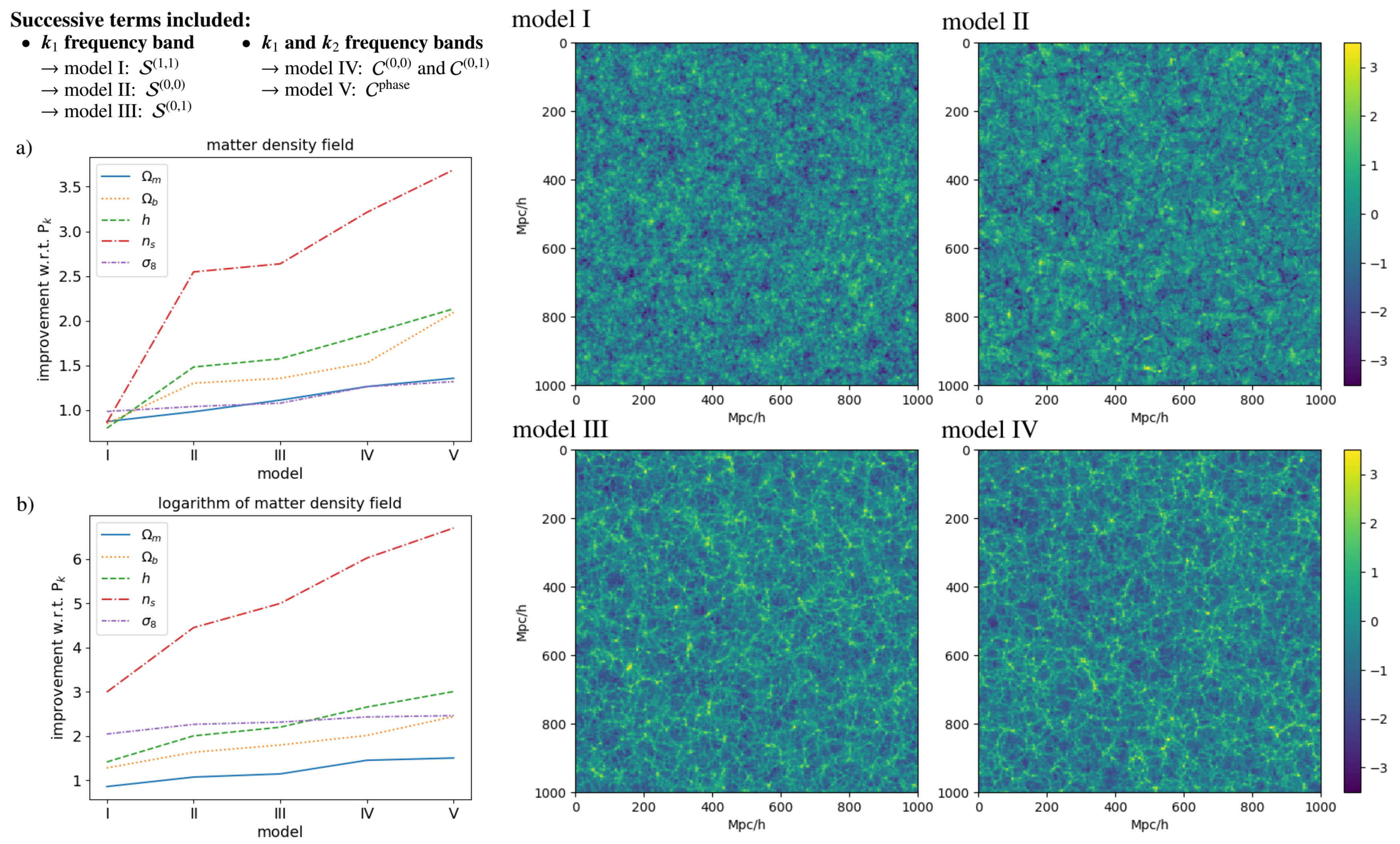}
\vspace{-0.5cm}
\end{center}
 \caption{Left: Improvement in the marginalized errors on cosmological parameters obtained with Fisher analysis using models I to V of WPH statistics, applied to (a) the projected matter density field $\rho$ and (b) its logarithm. In both cases, the results are normalized by those obtained with power spectrum statistics of the projected matter density field. Right: Syntheses of $\log(\rho)$ for models I to IV (see Fig.~\ref{FigResultSyntheses1}b for the corresponding model V synthesis). Both the Fisher analysis and the syntheses improve as the WPH statistics expand from model I to model V.}
\label{FigSbS_Results}
\end{figure*}

The results of these comparisons show that the syntheses from WPH statistics presented in this section perform remarkably well in reproducing the statistical properties of the logarithm of the Quijote LSS matter density field. Indeed, our syntheses reproduce the mean of the empirical power spectrum, its standard deviation, and its correlation within $5\%$, $10\%$, and $10\%$, respectively. They also reproduce the squeezed and flattened bispectra within $10\%$ and $20\%$, respectively, at spatial frequencies higher than $k=0.1~h{\rm Mpc}^{-1}$, and within $40\%$ below this spatial frequency.  The whole pixel PDF is also very well reproduced, including the tails down to 4 orders of magnitude below the peak.
There is no particular discrepancy between the synthetic and the Quijote distributions of the empirical power spectrum, as shown in Fig.~\ref{FigResultSyntheses2} at three arbitrary frequencies. 
Finally, the syntheses reproduce the three MFs to within $0.5\%$, $0.05\%$, and $0.02\%$, respectively.

Note that the syntheses do not accurately reproduce the equilateral bispectrum (Fig.~\ref{FigBkEqui}), although they do capture its general shape and changes of sign. This result may be related to the fact that equilateral bispectrum configurations correspond to correlations between three clearly separate frequencies, $\vec k_1, \vec k_2, \vec k_3$. Thus moments built from convolutions of only two wavelets cannot directly characterize them. Extending the WPH construction to characterize such couplings is left to future work.

Previous works have used convolutional neural networks, and especially GANs, to produce syntheses of astrophysical fields. For example, \cite{ramanah2020super} uses a GAN to emulate accurate high-resolution features from computationally cheaper low-resolution cosmological simulations. Similarly, in~\cite{aylor2019cleaning}, the authors use GANs to generate maps representing the interstellar medium. More recently, \cite{Tamosiunas:2020rvw} trained GANs to reproduce both weak lensing convergence maps and dark matter over-density fields. All these works assess the quality of the syntheses as in the present paper, i.e., by computing histograms, power spectra, bispectra, and Minkowski functionals. Our results are of similar quality to these earlier works.

However, because our method relies on explicit construction of statistics, it is not subject to the usual caveats of neural network methods. First of all, the WPH statistics can be physically interpreted (see Sec.~\ref{PartDiscussionVariousTerms}). Secondly, neural network methods must learn a large number of parameters (weights) from a huge training dataset, whereas we used only $30$ Quijote maps for our syntheses. Also, the use of GANs to generate new realizations of a given process can suffer from mode collapse, i.e., the omission of certain object classes in the generated images and loss of the associated statistical features~\citep{bau2019seeing}. These points underline the advantages of maximum entropy syntheses built with a suitably tailored statistical description such as the WPH moments.


\section{Physical interpretation of the WPH statistics}
\label{PartDiscussionVariousTerms}

This section discusses the physical meaning of the various WPH moments and their relation to other summary statistics. 
It complements the discussion of symmetries of Sec.~\ref{PartSymmetries}. See also Appendix~\ref{AppendixFinalModels} for the full specification of the WPH statistics.

To identify the physical properties encoded in the  WPH moments, we organize the moments into five categories. We define each category by selecting a set of $\left\{\vec \xi_1,p_1,\vec \xi_2,p_2 \right\}$ parameters. The first three categories contain moments that each have only a single spectral wavelet band (i.e., $\vec{\xi}_1 = \vec{\xi}_2$; see Sec.~\ref{PartAnalysisL1L2} below), while the other two contain moments describing a coupling between two wavelet bands of central frequencies $\vec{\xi}_1$ and $\vec{\xi}_2$ (see Sec.~\ref{PartAnalysisCoupling}).

To study these different categories of moments, we progressively include them in our analysis, and build five nested sets of statistics that we call model~I to model~V. While model I merely contains power spectrum information, model V corresponds to the WPH statistics used in Sec.~\ref{PartCosmoInfo} and~\ref{PartSyntheses}. Using each of these models, we compute the Fisher information for the Quijote simulated LSS matter density field $\rho$ and also for $\log(\rho)$, and perform statistical syntheses of $\log(\rho)$. 

\begin{figure*}[t]
\begin{subfigure}{.98\textwidth}
  \centering
  \includegraphics[width=1\linewidth]{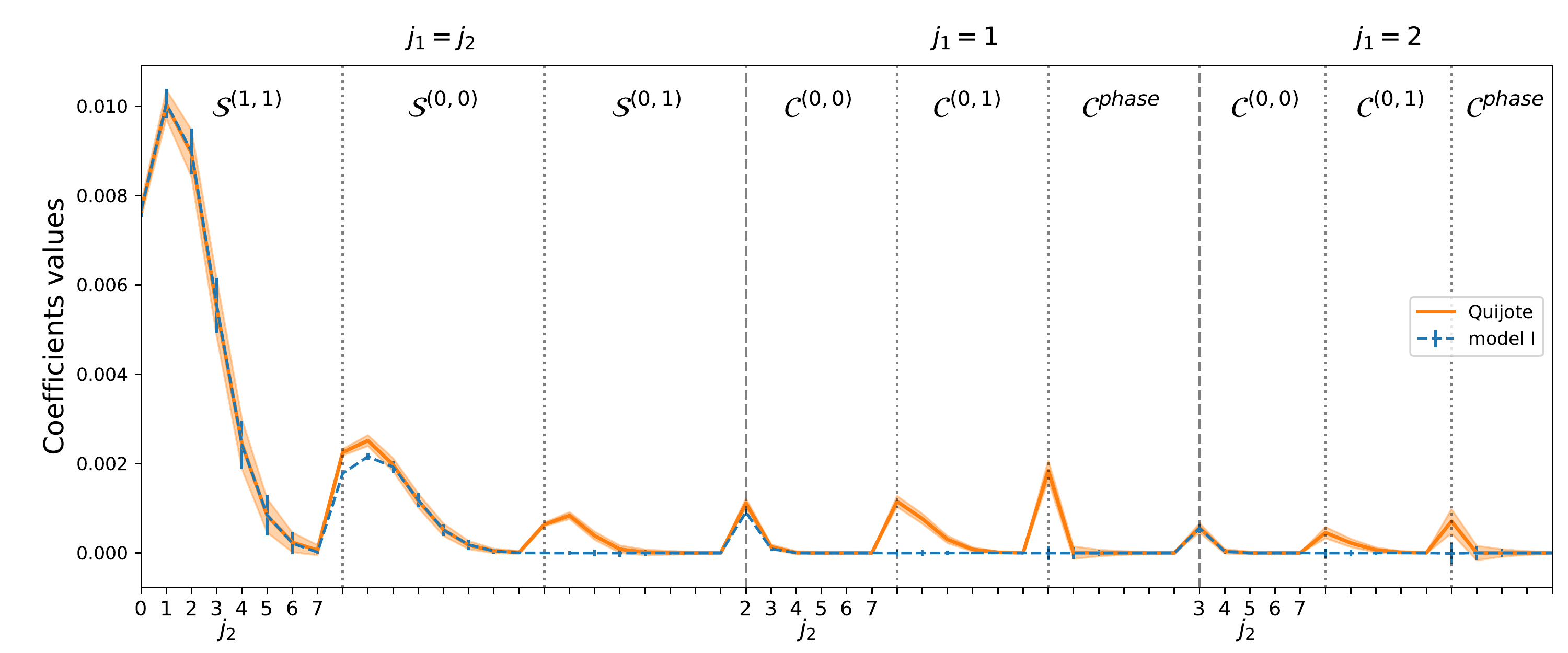}  
\end{subfigure}
\vspace{-0.16 cm}
\newline
\begin{subfigure}{.98\textwidth}
  \centering
  \includegraphics[width=1\linewidth]{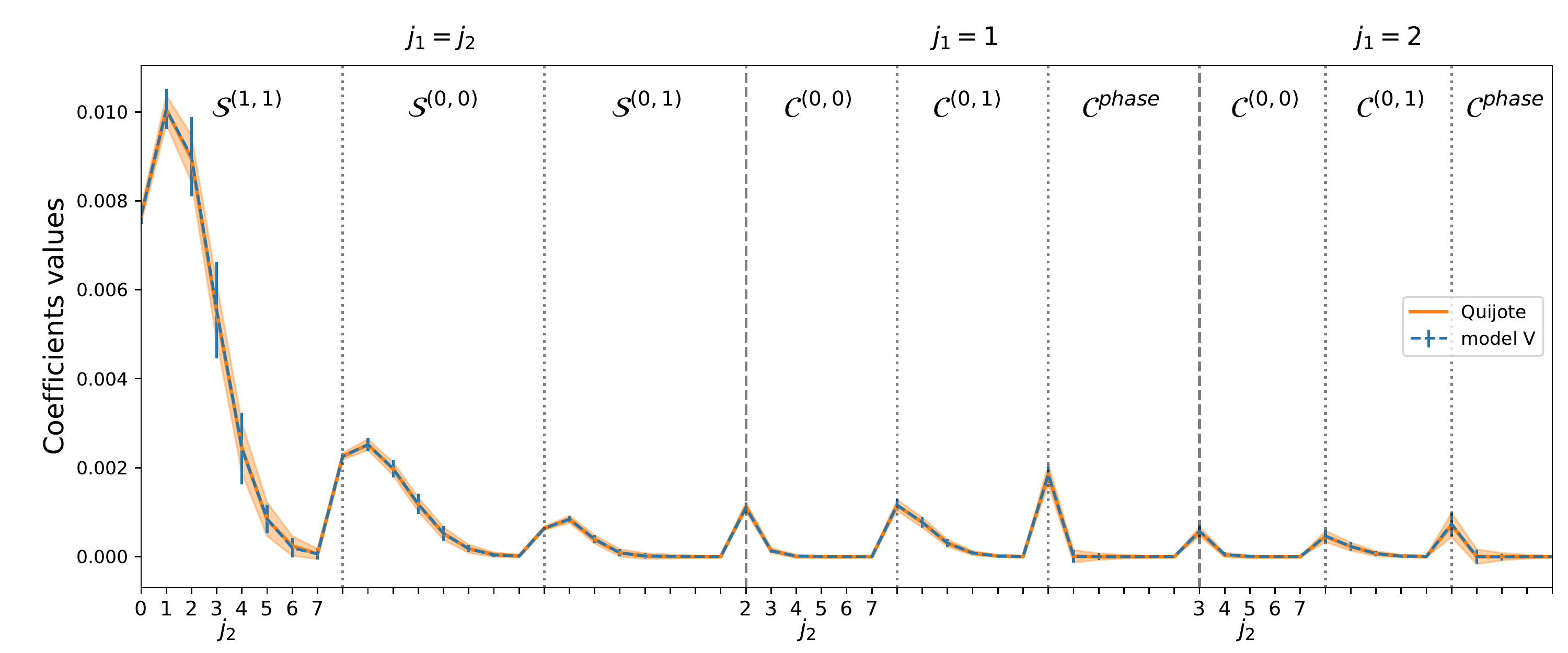}  
\end{subfigure}
\vspace{-0.25 cm}
 \caption{Subset of WPH moments, estimated on the logarithm of the matter density fields from Quijote simulations at fiducial cosmology (solid orange line, 15000 maps), and from model I (top) or model V (bottom) syntheses  (blue dashed line, 300 maps). The model I syntheses are a good approximation of a Gaussian field (see Sec.~\ref{PartAnalysisL1L2}). The model V corresponds to the WPH statistics used in Sec.~\ref{PartSyntheses}, which accordingly reproduce those of the Quijote simulation. The error bars, which characterize the variability of WPH moments from one map to another, have been multiplied by a factor of 3 for readability reasons. These moments are a subset of those used to estimate cosmological Fisher information in Sec.~\ref{PartCosmoInfo}. 
 For $\mathcal{S}$-type moments, all $j_1=j_2$ values are given. %
 $\mathcal{C}$-type moments are drawn for $j_1=1$ and $j_1=2$ only and for $j_1< j_2 \leq 7$.
 All plotted coefficients satisfy $\delta \ell =0$ and $n=0$, see App.~\ref{AppFisherWPHModel} for details.}
\label{FigPlotCoeffs}
\end{figure*}

\begin{figure}[t]
\begin{center}
\includegraphics[width = 0.55\textwidth]{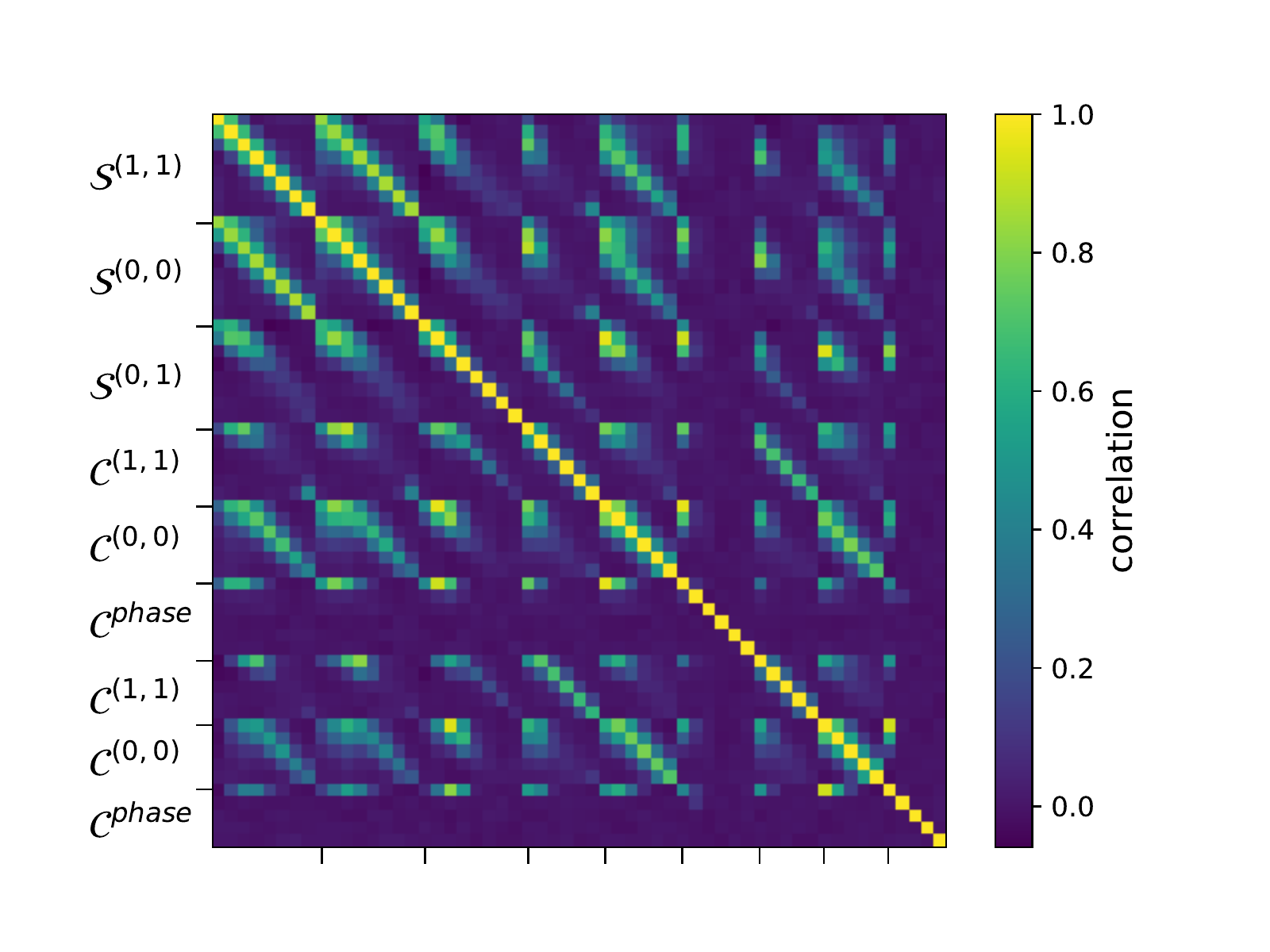}
\vspace{-1.3cm}
\end{center}
 \caption{Correlation matrix of a subset of WPH moments, estimated on the logarithm of the matter density field from Quijote simulations at fiducial cosmology (15000 maps). The WPH moments are those described in Fig.~\ref{FigPlotCoeffs}, with the same order, and the reader is referred to this figure for more details.}
\label{FigPlotCorr}
\end{figure}

\subsection{Terms related to a single wavelet frequency band}
\label{PartAnalysisL1L2}

Let us first consider WPH moments that describe a single spectral wavelet band.
Thus we start from the moments $\mathcal{C}_{\vec \xi_1,p_1,\vec \xi_2,p_2} (\vec{\tau})$ given in Eq.~\eqref{EqDefWPH_Cov}, and restrict to cases where $\vec{\xi}_1 = \vec{\xi}_2$. Hence, we focus on moments of the form
\begin{equation}
    \label{eq:defSsingle}
    \mathcal{S}^{(p_1,p_2)}_{\vec \xi_1}(\vec{\tau}) 
    =
    \mathcal{C}_{\vec \xi_1,p_1,\vec \xi_1,p_2} (\vec{\tau}) 
\end{equation}
and we consider three such terms obtained by taking the phase exponents $(p_1,p_2)$ equal to $(1,1)$, $(0,0)$, or $(0,1)$.
Explicitly, these are
\begin{align}   
\mathcal{S}^{(1,1)}_{\vec \xi_1}(\vec{\tau}) 
&=
\text{Cov} \left( \rho * \psi_{\vec \xi_1} (\vec x),\ \rho * \psi_{\vec \xi_1}  (\vec x + \vec \tau) \right),
\label{Eq11Single}
\\
\mathcal{S}^{(0,0)}_{\vec \xi_1}(\vec{\tau}) 
&=
\text{Cov} \left(  \left|\rho * \psi_{\vec \xi_1}(\vec x)\right|, \ \left|\rho * \psi_{\vec \xi_1}(\vec x + \vec \tau)\right| \right),
\label{Eq00Single}
\\
\mathcal{S}^{(0,1)}_{\vec \xi_1}(\vec{\tau}) 
&=
\text{Cov} \left(  \left|\rho * \psi_{\vec \xi_1}(\vec x)\right|,\ \rho * \psi_{\vec \xi_1}(\vec x + \vec \tau) \right).
\label{Eq01Single}
\end{align}
Note that $\mathcal{S}^{(1,1)}$ corresponds to the covariance of wavelet transforms presented in Sec.~\ref{PartCovWavelet}. 

\paragraph*{Model I: $\mathcal{S}^{(1,1)}$ moments; power spectrum.}

\begin{figure}[t]
\begin{center}
\includegraphics[width = 0.49\textwidth]{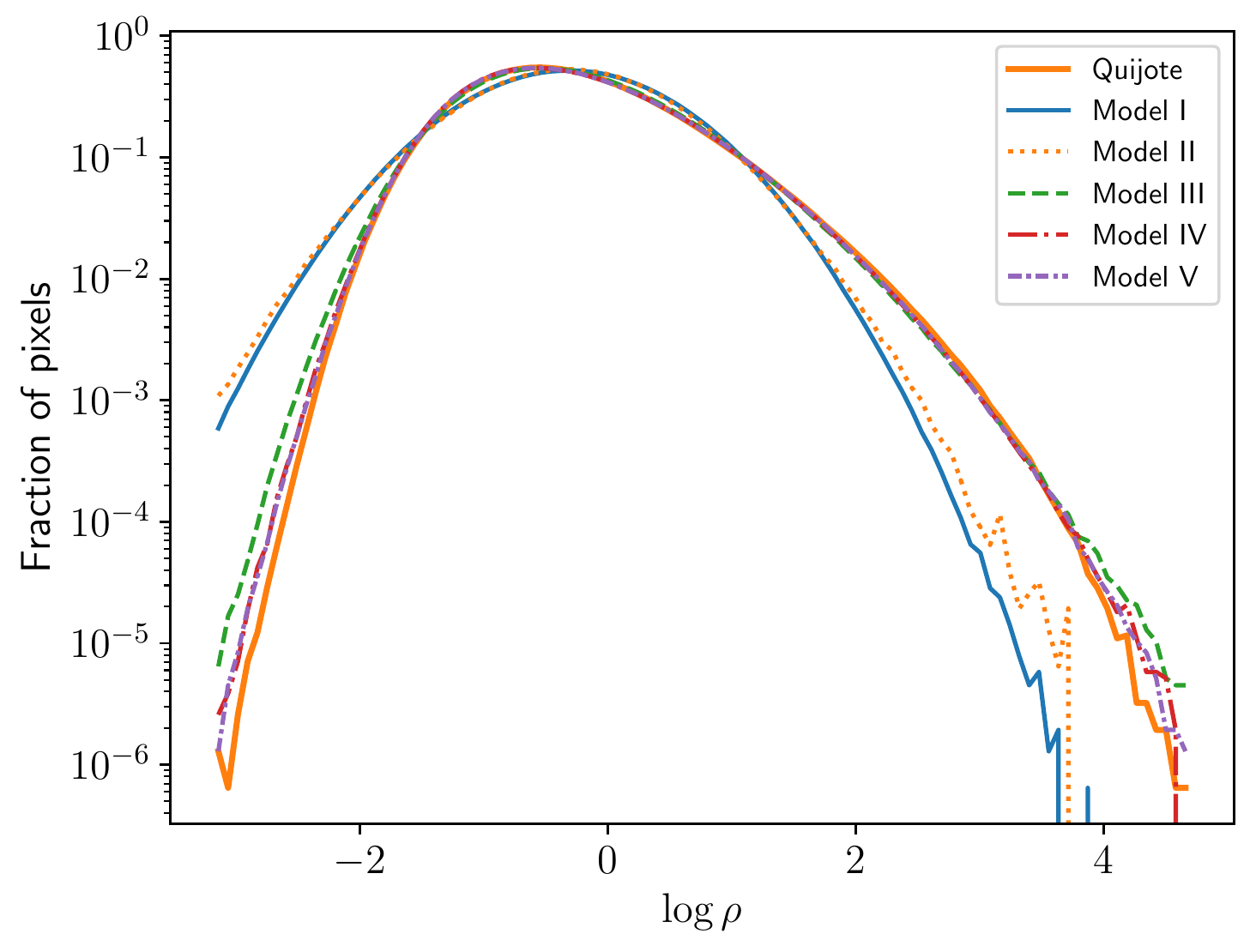}
\vspace{-1cm}
\end{center}
 \caption{Probability density function of statistical syntheses of $\log(\rho)$ as performed with models I to V, along with the PDF of the initial Quijote field. Each line is estimated from 300 maps, as in Fig.~\ref{FigResultSyntheses1}h.}
\label{FigSbS_Histo}
\end{figure}

As shown in Sec.~\ref{PartCovWavelet}, the $\mathcal{S}^{(1,1)}$ moments, which form model I WPH statistics, only depend on the power spectrum of $\rho$. Thus we see that syntheses from model I generate fields close to Gaussian (see Fig.~\ref{FigSbS_Results}). These fields have in particular a Gaussian PDF (see Fig.~\ref{FigSbS_Histo}) and the ensemble mean of their bispectra is null. 

With these moments we can also see the impact on the spectral resolution of the number of relative spatial translations $\vec{\tau}$ used (see Eq.~\ref{eqDefTauNTheta} of Appendix~\ref{AppendixFinalModels}). Figure~\ref{FigPkDnImprovement} illustrates this property by showing the power spectrum of the model I syntheses for two cases: The imprints of the wavelet spectral bands are clearly visible when no translations are used (i.e., $\vec{\tau}=0$ only, which corresponds to $\Delta_n=0$), while they are reduced when $\Delta_n = 2$ (i.e., two nonzero translations $\vec{\tau}$ in each direction). We obtained similar results for the Fisher analysis of cosmological parameters. Indeed, the amount of information on cosmological parameters carried by model I WPH statistics increases with the number of retained values of $\vec{\tau}$, eventually converging to the information contained in the standard power spectrum.

\begin{figure}[t]
\begin{center}
\includegraphics[width = 0.49\textwidth]{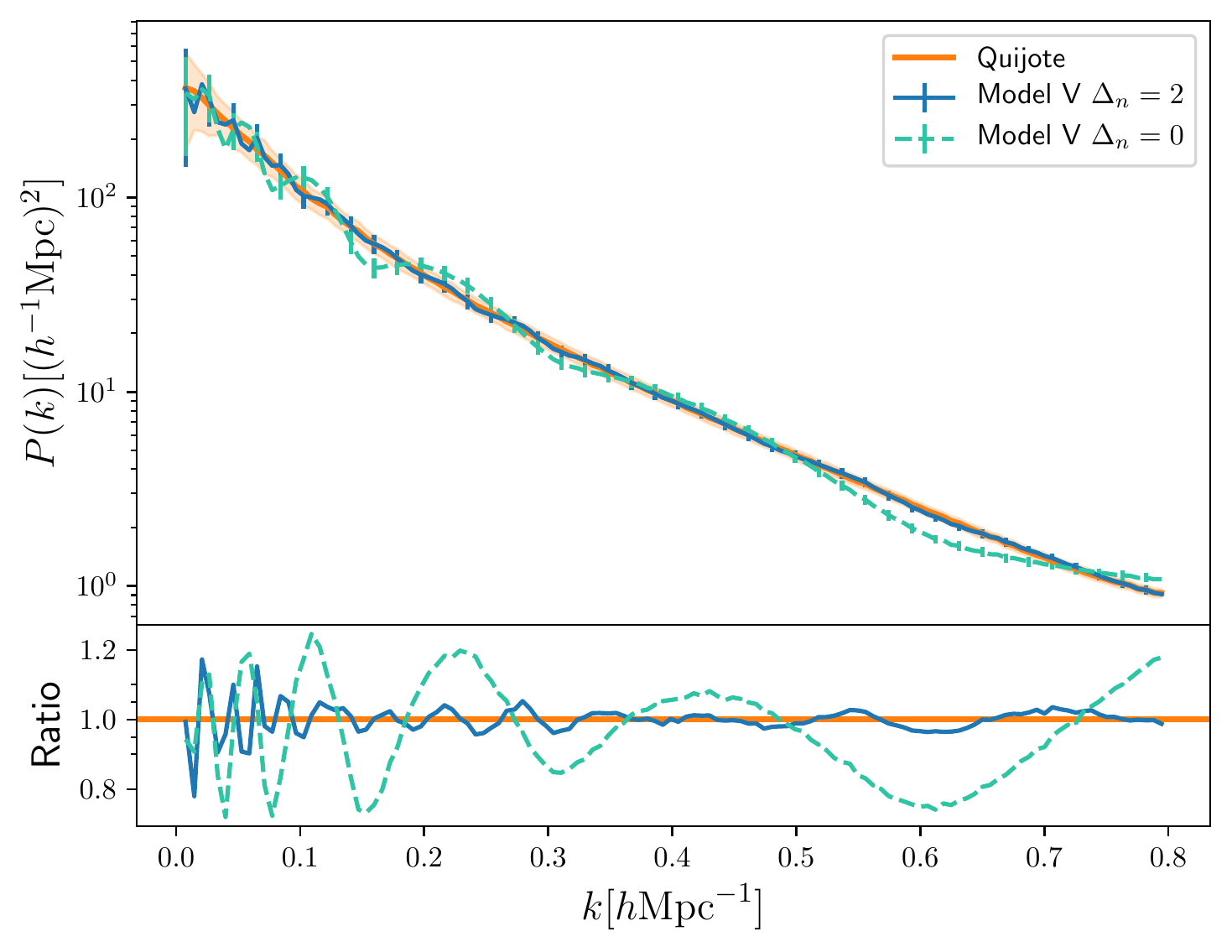}
\vspace{-1cm}
\end{center}
 \caption{Improvement of the power spectrum of model I syntheses of $\log(\rho)$ as the number $\Delta_n$ of spatial translations in each direction is increased from 0 to 2. The imprint of the wavelet spectral bands are clearly visible at $\Delta_n = 0$ (i.e., no spatial translation) and are greatly reduced at $\Delta_n = 2$. The error bars correspond to the realization-per-realization dispersion.}
\label{FigPkDnImprovement}
\end{figure}

\begin{figure*}[t]
\begin{center}
\includegraphics[width = 0.95\textwidth]{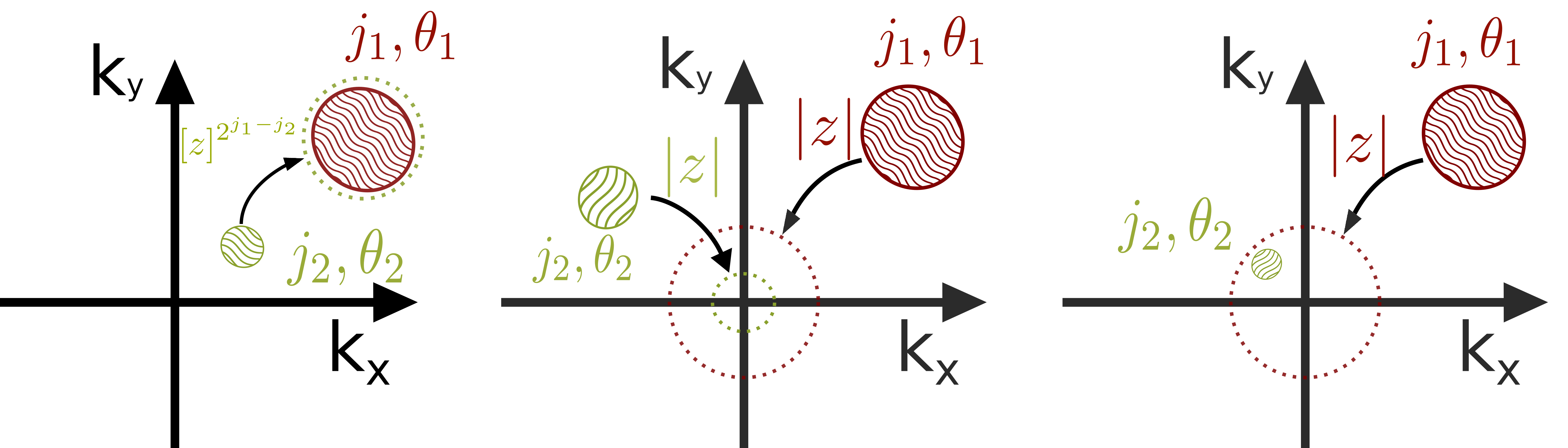}
\vspace{-0.5cm}
\end{center}
 \caption{Coupling terms depicted in Fourier space for (from left to right) the $\mathcal{C}^\text{phase}$, $\mathcal{C}^{(0,0)}$, and $\mathcal{C}^{(0,1)}$ types of coupling. In each case the frequency support of the filtered fields (filled circles) are modified by the phase harmonic operator to share common frequencies (dotted circles). Note that this figure is for explanatory purposes only, since in practice the spectral support of a filtered field and its modulus overlap.}
\label{FigCouplingFourier}
\end{figure*}

\paragraph*{Model II: $\mathcal{S}^{(0,0)}$ moments and sparsity.}

We build model II by adding the moments $\mathcal{S}^{(0,0)}$ to model I. These moments allow us to quantify the ratio between the $\mathbf{L}^1$ and $\mathbf{L}^2$ norms of the wavelet transform of $\rho$. This ratio characterizes the sparsity of the field in the wavelet basis \citep[see][for a more detailed discussion]{zhang2019maximum}. In the Quijote LSS simulations, these coefficients indicate that the small scales are sparser than the large ones, and as the scale increases the sparsity converges to that of a Gaussian field, see Fig.~\ref{FigPlotCoeffs}. We expect such a result, since the LSS density fields are expected to become Gaussian at scales larger than 100 Mpc$/ h$ (about 25 pixels in our maps).

The amount of Fisher information with respect to the cosmological parameters is significantly higher in model II than in model I, as shown by Fig.~\ref{FigSbS_Results}-a,b. Similarly to model I, increasing the number of spatial shifts $\vec{\tau}$ in model II substantially improves the Fisher information with respect to the cosmological parameters.

The $\mathcal{S}^{(0,0)}$ moments do not substantially improve the synthesis results, as can be seen in the maps in Fig.~\ref{FigSbS_Results}. 

\paragraph*{Model III: $\mathcal{S}^{(0,1)}$ moments and first structures.}

To build the third model of WPH statistics, we add the $\mathcal{S}^{(0,1)}$ moments to model II. These moments measure a covariance between wavelet coefficients that undergo two different operations (namely, modulus and identity). Hence they mainly describe couplings between different spatial frequencies within a single wavelet band, see Sec.~\ref{PartWPH_Concept}. Thus, they probe the impact of the interaction between neighboring scales.

The $\mathcal{S}^{(0,1)}$ moments significantly improve the syntheses generated. The model III map in Fig.~\ref{FigSbS_Results} illustrates this, where the familiar foamy structure of the LSS is recognized. In Fig.~\ref{FigSbS_Histo} we see that model III (unlike models I and II) reproduces the main shape of the PDF of the Quijote simulations. Similarly, the model III syntheses reproduce the Minkowski functionals and the flattened bispectrum, but not the squeezed bispectrum. This last result is due to the fact that squeezed bispectrum triangles characterize the joint information between Fourier modes of wave-vectors of very different sizes.

In spite of these results, model III makes only minor improvements to the Fisher information regarding cosmological parameters. This is especially noticeable for the LSS matter density field (Fig.~\ref{FigSbS_Results}a), and less so for its logarithm (Fig.~\ref{FigSbS_Results}b). These results emphasize that although the interactions between nearby scales are important statistics for qualitatively reproducing the web structure of the LSS, they do not seem efficient for discriminating between different cosmological parameter values. 

\subsection{Couplings between wavelet frequency bands}
\label{PartAnalysisCoupling}

The second type of moment that we consider characterizes couplings between two wavelet spectral bands with central spatial frequencies $\vec{\xi}_1$ and $\vec{\xi}_2$. We consider three kinds of such couplings based on the moments $\mathcal{C}_{\vec \xi_1,p_1,\vec \xi_2,p2}$, with different values of $p_i$. 
In each case, the two phase harmonics of wavelet transforms contain common spatial frequencies of oscillation.
For each term, we set $ \xi_2 \leq \xi_1 $ (corresponding to spatial scales $2^{j_2} \geq 2^{j_1}$), without loss of generality. 
Similarly to the definition of Eq.~(\ref{eq:defSsingle}), we define
\begin{equation}
    \label{eq:defSdouble} 
    \mathcal{C}^{(p_1,p_2)}_{\vec \xi_1,\vec \xi_2} (\vec \tau) 
    = \mathcal{C}_{\vec \xi_1,p_1,\vec \xi_2,p2}(\vec \tau).
\end{equation}
For the first two types of moment, we consider $(p_1,p_2)=(0,0)$ and  $(0,1)$.  Explicitly, those are
\begin{align}
    \mathcal{C}^{(0,0)}_{\vec \xi_1,\vec \xi_2} (\vec \tau)
    &= \text{Cov} \left( 
    \left| \rho * \psi_{\vec \xi_1}(\vec x)\right|,
    \left| \rho * \psi_{\vec \xi_2} (\vec x + \vec\tau)\right|
    \right)
\label{EqCoupling00}
\\
\mathcal{C}^{(0,1)}_{\vec \xi_1,\vec \xi_2} (\vec \tau) 
&= 
\text{Cov} \left( 
    \left| \rho * \psi_{\vec \xi_1}(\vec x) \right|,
    \rho * \psi_{\vec \xi_2} (\vec x + \vec \tau)
    \right).
\label{EqCoupling01}
\end{align}
For $\vec{\xi}_1 = \vec{\xi}_2$, the $\mathcal{C}^{(0,0)}$ and $\mathcal{C}^{(0,1)}$ moments are identical to $\mathcal{S}^{(0,0)}$ and $\mathcal{S}^{(0,1)}$ defined in the previous section. 
Note that $\mathcal{C}^{(0,1)}$ is not symmetric under exchange of $\vec \xi_1$ and $\vec \xi_2$, but it is negligibly small when $\xi_2 > \xi_1$, as discussed below.

These moments can be interpreted as follows. First, $\mathcal{C}^{(0,0)}$ quantifies the correlation between local levels of oscillations at $\vec \xi_1$ and $\vec \xi_2$ spatial frequencies. Then, $\mathcal{C}^{(0,1)}$ evaluates the correlation between the amplitude of the local level of oscillation at the $\vec \xi_1$ frequency and the oscillation at the $\vec \xi_2$ frequency. For this second moment, since $\rho * \psi_{\vec \xi_1}$ is filtered at a $2^{j_1}$ wavelength, it is clear that the correlation of its amplitude with a $\rho * \psi_{\vec \xi_2}$ term gives a negligible result if this second convolution oscillates at a characteristic scale $2^{j_2} < 2^{j_1}$.

\begin{figure*}[t]
\begin{subfigure}{.48\textwidth}
\begin{center}
\includegraphics[width = 0.99\textwidth]{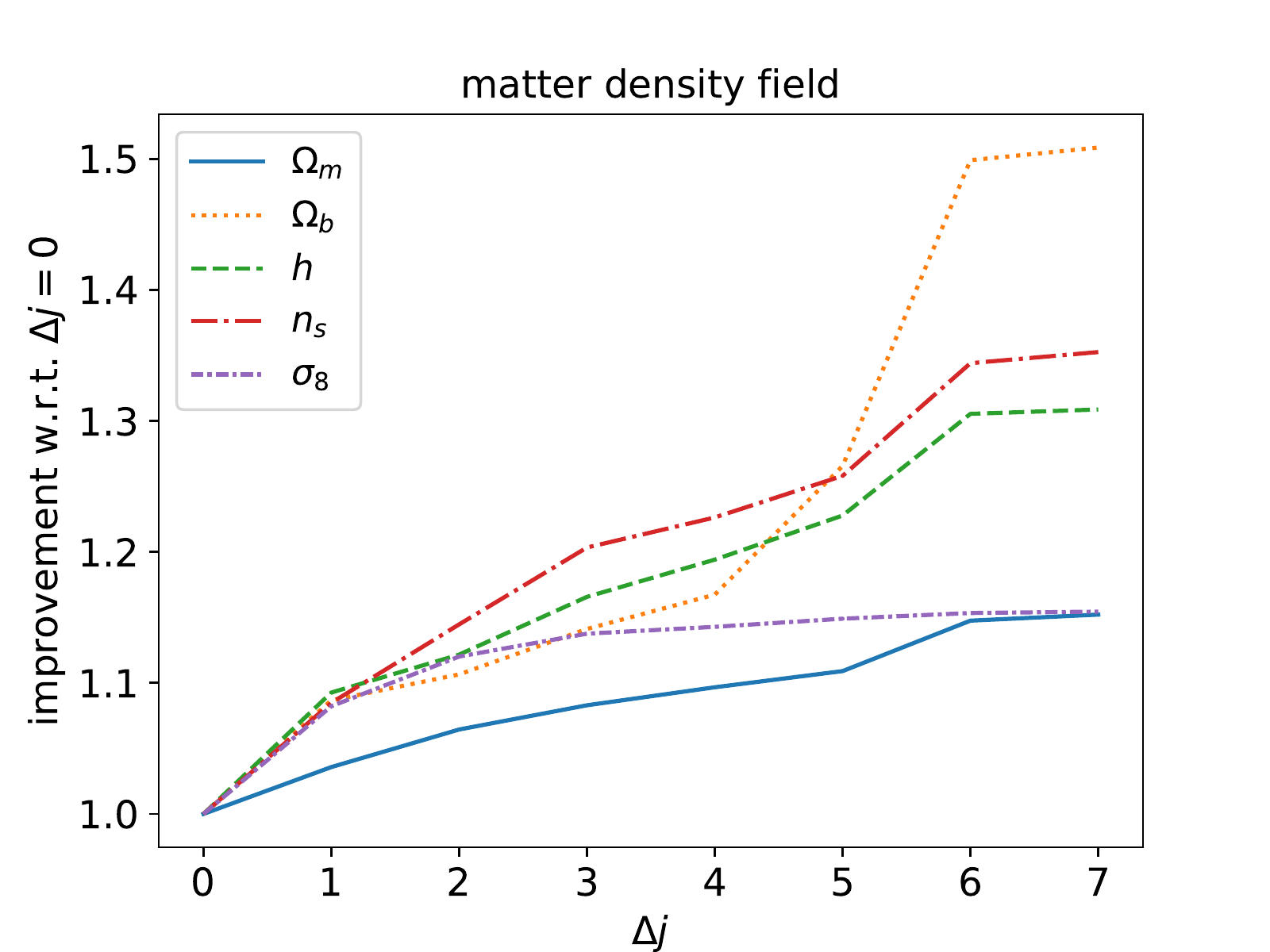}
\vspace{-0.15cm}
\end{center}
\end{subfigure}
\begin{subfigure}{.48\textwidth}
\begin{center}
\includegraphics[width = 0.99\textwidth]{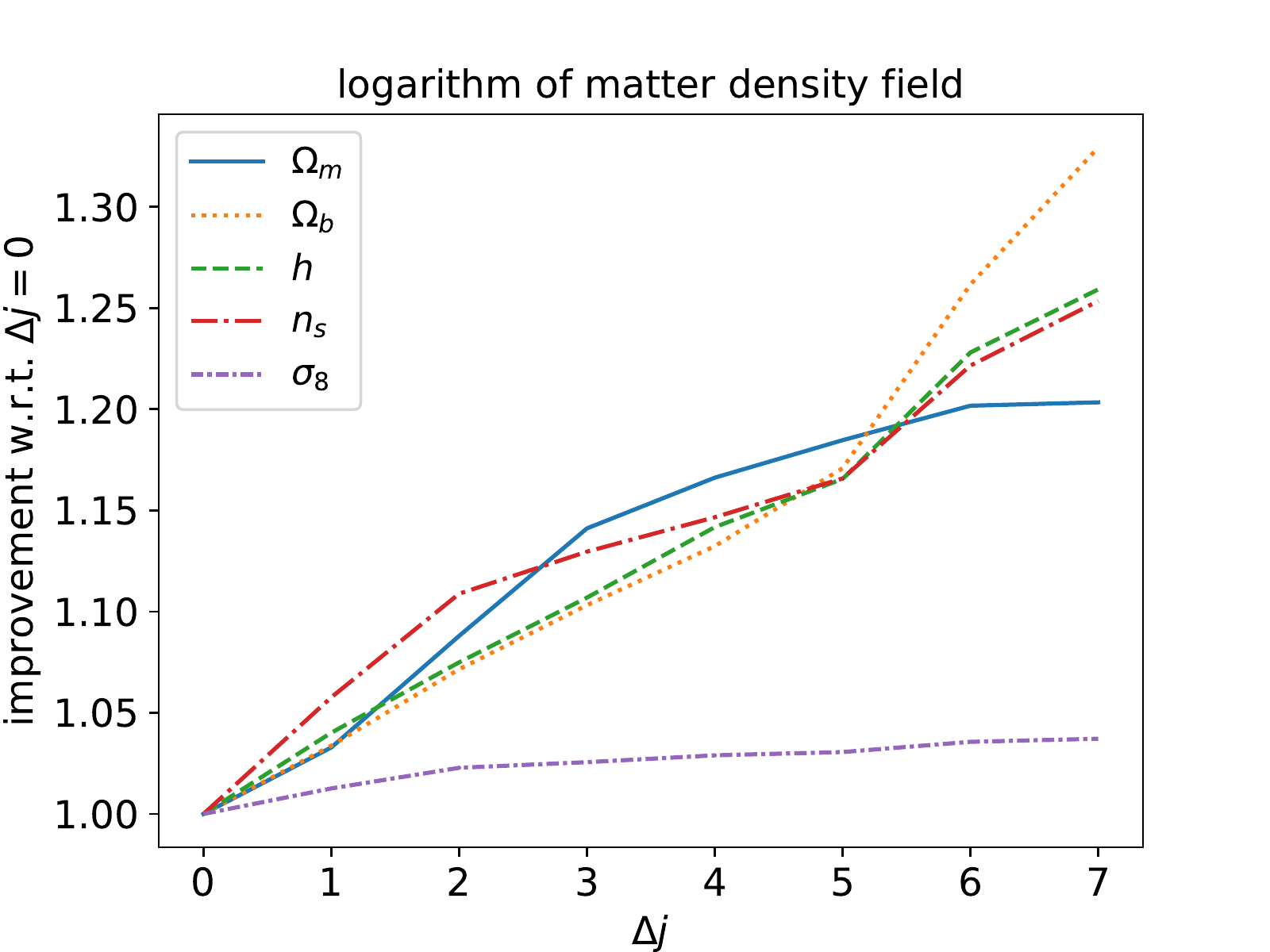}
\vspace{-0.15cm}
\end{center}
\end{subfigure}
 \caption{Improvements of the marginalized errors on cosmological parameters obtained with Fisher analysis for model V of WPH statistics with increasing $\Delta_j$ values, for the LSS matter density field (left) and its logarithm (right). Errors are normalized by those obtained with $\Delta_j=0$. 
 When limited to $\Delta_j=0$,
 model V characterizes only local couplings and corresponds to model III.
 }
\label{FigDeltajImprovement}
\end{figure*}

Our last coupling uses $(p_1, p_2)=(1, \xi_1 / \xi_2)$.  We define
\begin{eqnarray}
    \label{EqPhaseCoupling}
    \mathcal{C}^\text{phase}_{\vec \xi_1,\vec \xi_2}(\vec{\tau}) 
    & = & \mathcal{C}_{\vec \xi_1,1,\vec \xi_2,\xi_1/\xi_2}(\vec{\tau})
    \nonumber \\
    & = & \text{Cov} \left(  
        \rho * \psi_{\vec \xi_1}(\vec{x}),
        \left[\rho * \psi_{\vec \xi_2}(\vec{x} + \vec{\tau})\right]^{\xi_1/\xi_2} 
        \right).
\end{eqnarray}
For $\vec{\xi}_1 = \vec{\xi}_2$, this moment is equivalent to $\mathcal{S}^{(1,1)}$. The computation of such WPH moments is illustrated in Fig.~\ref{FigConceptWPH} and was discussed in Sec.~\ref{PartWPH_Concept}. Computed from fields filtered at different scales which are made synchronous, such terms are designed to characterize the relative phase shift between different scales. 

These three coupling terms can also be understood from a Fourier space point of view. For each of them, Fig.~\ref{FigCouplingFourier} illustrates how the phase harmonics operator modifies the spectral support of the $\rho$ field convolved with the $\vec{\xi}_1$ and $\vec{\xi}_2$ wavelets. The three coupling terms correspond to different ways of achieving a (possibly partial) spectral overlap by band-passing followed by the phase operation.


\paragraph*{Models IV \& V: couplings between different wavelet bands.}

We define model IV by adding the $\mathcal{C}^{(0,0)}$ and $\mathcal{C}^{(0,1)}$ moments to model III. Finally, we define Model V by adding the $\mathcal{C}^{\text{phase}}$ moments. Note that these models include terms for which $\vec{\xi}_1$ and $\vec{\xi}_2$ have the same norm but different orientations. Model V corresponds to the statistical descriptions used in Sec.~\ref{PartCosmoInfo} and~\ref{PartSyntheses}.

The syntheses generated from model IV are significantly better than those of model III. This model reproduces the squeezed bispectrum triangles, as well as the tails of the PDF (with results that are better than model III by a factor close to 5). This result underlines that $\mathcal{C}^{(0,0)}$ and $\mathcal{C}^{(0,1)}$ moments are related to couplings between scales that are far apart. In particular, the PDF result exhibits how the precise distribution of peaks of the LSS seems to be related to couplings between different scales that sum up together in a coherent way. By contrast, model V does not significantly improve the syntheses compared to model~IV.

For cosmological parameter inference, Fig.~\ref{FigSbS_Results} shows how both models IV and V noticeably improve the forecast errors. These improvements are generally significantly larger than those obtained with the local coupling added in model III. It is interesting to see that while the $\mathcal{C}^{\text{phase}}$ moments play only a minor role in reproducing the standard statistics in syntheses, they do contain a substantial amount of information about cosmological parameters. This result
could indicate that the $\mathcal{C}^{\text{phase}}$ moments are not directly related to the summary statistics we used to validate the syntheses. 

\paragraph*{Importance of the ratio between coupled scales.}

For a given set of WPH statistics, the value $\Delta_j = j_\text{max} - j_\text{min}$ quantifies the maximum scale difference being characterized by the moments. Indeed, the maximum ratio between such scales is $2^{j_\text{max}}/2^{j_\text{min}} = 2^{\Delta_j}$. Similarly, the ratio between the norm of the more distant spatial frequencies that are coupled is $\xi_1/\xi_2 = 2^{\Delta_j}$. 

For nonlinear physical processes, this parameter is of major importance. Indeed, nonlinearity implies a statistical interaction between different scales. The more nonlinear a given process is, the stronger we expect distant scales to be coupled~\citep{bruna2015intermittent}. We also expect different nonlinear couplings to have distinct signatures in the way scales decouple from one another when the ratio between scales increases.

\begin{figure}[t]
\begin{center}
\includegraphics[width = 0.47\textwidth]{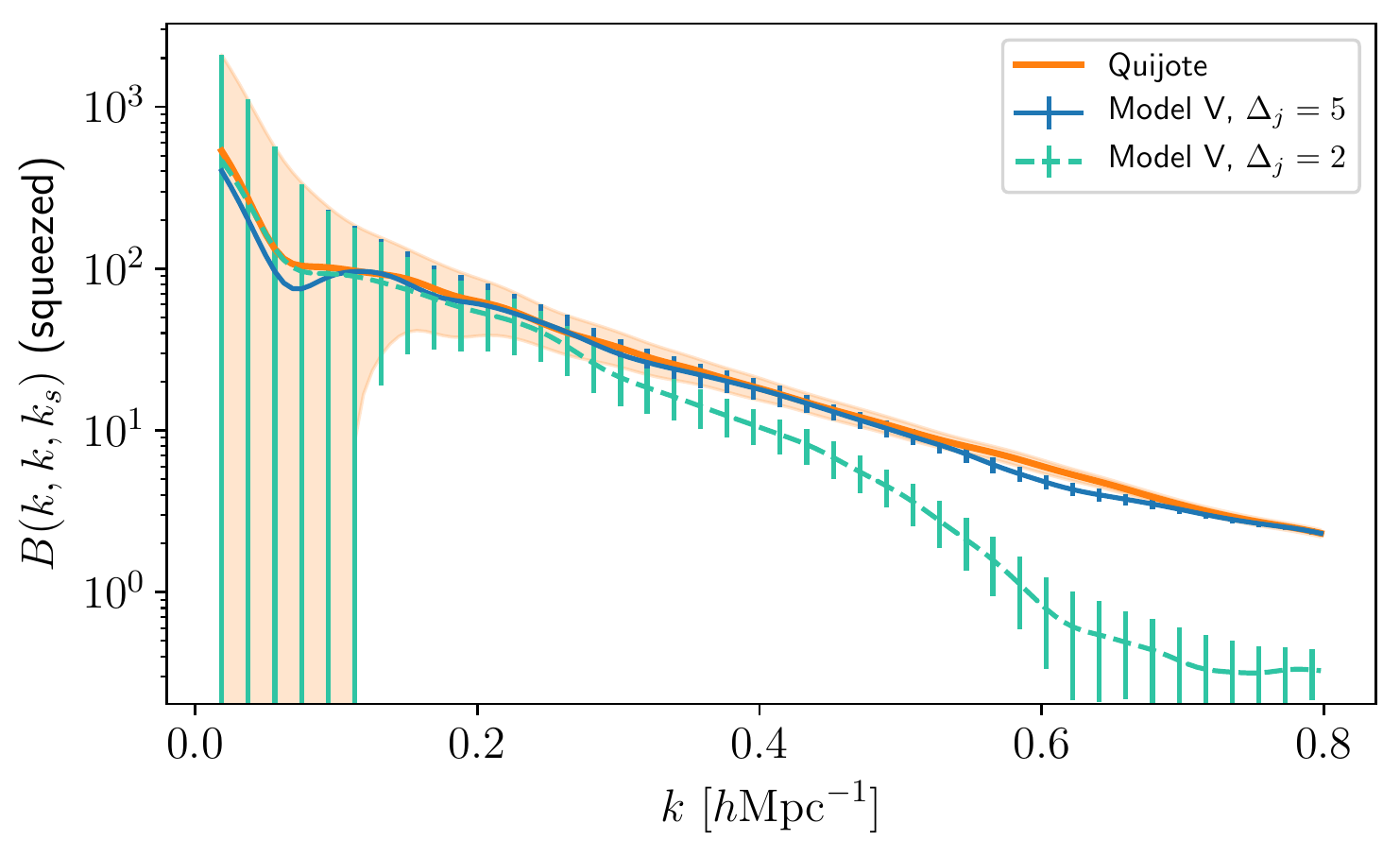}
\vspace{-0.6cm}
\end{center}
 \caption{Squeezed-triangle bispectrum of $\log(\rho)$ of the Quijote simulation (orange) and its syntheses from model V of WPH statistics with $\Delta j =2$ (green) and $\Delta j = 5$ (blue). The $\Delta j =2$ synthesis does not characterize the couplings between distant scales of the LSS that are needed to reproduce this bispectrum. The error bars correspond to the realization-per-realization dispersion.}
\label{FigDeltajImprovement2}
\end{figure}

Let us illustrate the importance of $\Delta_j$. For cosmological parameter inference, one can compute how the Fisher forecast errors evolve for model V as we increase $\Delta_j$ from 0 (which corresponds to model III). As shown in Fig.~\ref{FigDeltajImprovement}, we find that significant information is contained in the coupling between different scales. The Fisher results keep improving even when we add terms with $\Delta_j \geq 5$, representing coupling between scales that are very far apart (e.g., 2 and 64 pixels). This especially contrasts with the minor improvements brought about by the inclusion of couplings between nearby scales.

The increase of the Fisher information with $\Delta_j$ is different for each cosmological parameter. In particular, this improvement is only modest for $\Omega_m$, and especially small for $\sigma_8$ above $\Delta_j =2$. Note that $\Omega_m$ and $\sigma_8$ are the two parameters for which the WPH statistics do not characterize much more information than the power spectrum (see Table~\ref{TableFisherResultsMain}). This indicates that these particular parameters do not especially impact the way distant scales couple. This result seems rather natural for $\sigma_8$, since it is merely a normalization of the matter fluctuation power spectrum. 

The importance of $\Delta_j$ also appears for the syntheses. Even though the cosmic web structure visually appears with a model where $\Delta_j = 0$ (see the model III map in Fig.~\ref{FigSbS_Results}), larger values of $\Delta_j$ are needed to properly reproduce the tails of the PDF, which for instance characterize the distribution of peaks of the LSS. This can be seen by carefully comparing tails of the model III, IV, and V results in Fig.~\ref{FigSbS_Histo}. Similarly, larger values of $\Delta_j$ must be included to reproduce the squeezed triangle bispectrum of the Quijote LSS field (see Fig.~\ref{FigDeltajImprovement2}). This result is, however, completely to be expected, since those triangles characterize couplings between very different scales.


\section{Conclusion}
\label{PartConclusion}

In this paper, we introduced low-dimensional Wavelet Phase Harmonics (WPH) statistics to analyze and synthesize two-dimensional matter density fields from the Quijote LSS simulations. We built the WPH statistics from WPH moments, which were recently introduced in data science. The WPH moments correspond to the covariance of wavelet coefficients whose spatial frequencies are made synchronous by means of a nonlinear operator called the phase harmonic operator. 

The main result of this paper is the construction of specific low-dimensional WPH statistics that achieve state-of-the-art results for both capturing cosmological information and producing statistical syntheses. We obtained these results by computing the Fisher information of these statistics with respect to five cosmological parameters, and by producing maximum-entropy syntheses that were validated by means of classic summary statistics. To our knowledge, it is the first time that state-of-the-art results have been obtained for both of these tasks from the same statistical description. We also illustrated the interpretability of WPH statistics by discussing the types of information described by distinct subsets of WPH moments.

In this paper, we applied the WPH statistics to the projected LSS matter density field. However, their construction is not specific to this process, and they can therefore be used to study other non-Gaussian physical fields. A natural extension is to apply this method to three-dimensional fields, which is left for future work.
This would allow a direct comparison with results obtained from other summary statistics or from machine learning methods. 

The quality of the statistical syntheses produced in this paper validates the relevance of using WPH statistics to study the LSS. Moreover, this work shows that WPH statistics can serve as a generative model for non-Gaussian fields. They could be used to generate mock syntheses, or to perform data augmentation for machine learning purposes. The syntheses can be performed from a limited training set (here, we used 30 Quijote maps), and therefore we could train our model directly on observational data.

\section*{Acknowledgments}

We thank D. Spergel, B. M\'enard, and B. Wandelt, for fruitful discussions, as well as F. Levrier and B. R\'egaldo-Saint Blancard for their comments on the draft. We also thank G. P. Collins for his valuable assistance in the preparation of this document, as well as the anonymous referee, whose comments helped to clarify the paper. EA thanks the Flatiron Institute of the Simons Foundation for its generous hospitality during the preparation of this work. FVN acknowledges funding from the WFIRST program through NNG26PJ30C and  NNN12AA01C. SM also acknowledges support from the PRAIRIE 3IA Institute of the French ANR-19-P3IA-0001 program.

\appendix


\section{Mathematical specifications}
\label{AppMathSpec}
\subsection{Bump steerable mother wavelet}
\label{AppendixBSWavelets}

The multiscale bump steerable wavelets $\psi_{j,\ell}$ used for the wavelet transform described in Sec.~\ref{PartWavelet} are built from a mother wavelet $\psi$. Following~\cite{mallat2018phase}, which first introduced bump steerable wavelets, we define $\psi$ via its Fourier transform $\hat \psi(\vec k)$ as
\begin{multline}
    \hat \psi (\vec k) = c \cdot \exp \left(\frac{-(||\vec k|| - \xi_0)^2}{\xi_0^2 - (||\vec k||-\xi_0)^2}\right) \cdot \mathbf{1}_{[0, 2\xi_0]}(||\vec k||)  \\
    \times \cos^{L/2-1} (\arg (\vec k)) \cdot \mathbf{1}_{[0, \pi/2]}(|\arg (\vec k)|),
\end{multline}
 where $(\xi_0, 0)$ is the central frequency of the wavelet, $c$ is a normalization constant, $L$ is the number of angles used in the multiscale wavelet family, and $\mathbf{1}_A(x)$ is the indicator function that returns 1 if $x\in A$ and 0 otherwise.
 Following \cite{zhang2019maximum}, we use $\xi_0 = 1.7\pi$, and $c = 1.29^{-1} 2^{L/2 -1} \frac{(L/2 -1)!}{\sqrt{(L/2)(L-2)!}}$ with $L=16$.

\subsection{Scaling functions}
\label{AppendixScalingFunction}

We also consider convolutions of the field $\rho$ with a family of low-pass filters $\varphi_{j}$ called \textit{scaling functions}. These low-pass filters are built from an initial Gaussian window $\varphi$ defined as
 \begin{equation}
     \hat \varphi (\vec k) = \exp \left(-\frac{||\vec k||^2}{2 \sigma^2} \right).
 \end{equation}
The individual $\varphi_{j}$ are obtained from $\varphi$  by dilations of $2^j$:
\begin{align}
\varphi_{j} (\vec{x}) &= 2^{-j} \varphi\left(2^{-j}\vec{x}\right), \\
\hat \varphi_{j} (\vec{k}) &= 2^j \hat \varphi\left(2^j \vec{k}\right).
\end{align}
Again following \cite{zhang2019maximum}, we use $\sigma = 0.248\times 2^{-0.55}\xi_0$.

\subsection{Bispectrum estimates}
\label{AppendixBSTriangle}

For our bispectrum computations, we have adapted the method described in \cite{JunGRacineBartjan} by
which a smoothed isotropic bi\-spectrum is estimated as a 3-point correlation between three filtered versions of the field.
Specifically, we use isotropic filters $h_i$ which select only frequencies $\vec{k}$ such that $\|\vec{k}\| = k_i$, and we define $\rho_i = \rho * h_i$. The bispectrum $B(k_1, k_2, k_3)$ is then estimated from $\langle\rho_1(\vec{x})\rho_2(\vec{x})\rho_3(\vec{x})\rangle$.
In particular, we use
\begin{equation}
\label{EqFilters}
h_i(\vec{k}) = \frac{1}{\sigma\sqrt{2\pi}}\exp\left[\frac{-\big\|\vec k - \vec k_i\big\|^2}{2 \sigma^2}\right].
\end{equation}

We define $k_N = 1/256\ h \mathrm{Mpc}^{-1}$ as a reference value for wavenumbers, and we use $\sigma = 4 k_N$ in Eq.~\eqref{EqFilters} for the bispectrum computations used to validate the syntheses in Sec.~\ref{PartSyntheses}.

In Sec.~\ref{PartCosmoInfo}, we use $\sigma = 2 k_N$, and we specify the following set of bispectrum statistics to compute Fisher information about cosmological parameters:
\begin{itemize}
\item All flattened triangle configurations $B(k,k/2,k/2)$, with $k=(2n+1)k_N$ for $n$ between 1 and 62.
\item All equilateral triangle configurations $B(k,k,k)$, with $k=(2n+1)k_N$ for $n$ between 1 and 62.
\item All squeezed triangle configurations $B(k,k,k_s)$, with $k=(2n+1)k_N$ for $n$ between 1 and 62, and $k_s = 4 k_N$.
\end{itemize}
This set contains 62 triangles of each type, for a total of 186 bispectrum terms.

\subsection{Minkowski functionals}
\label{AppMF}
We use three Minkowski functionals ($V_0$, $V_1$, $V_2$) to assess the quality of the syntheses in section \ref{sec:resultSynth}.
Given a threshold $\nu$, these are respectively the area, the perimeter, and the genus defined by the threshold. More precisely, for a field $I(\vec{x})$ defined on an area $A_{\text{tot}}$, let us define 
$\Gamma_{<\nu} = \{ \vec{x} : I(\vec{x}) < \nu\}$ 
and similarly for $\Gamma_{\geq\nu}$ and $\Gamma_{>\nu}$.
Let $A_\nu$ be the area of $\Gamma_{\geq\nu}$, $S_\nu$ its perimeter, and $C_{<\nu}$ ($C_{>\nu}$) the number of connected components of $\Gamma_{<\nu}$ ($\Gamma_{>\nu}$). Then:
\begin{align}
V_0(\nu) = \frac{A_\nu}{A_{\text{tot}}}, ~~ \ V_1(\nu) = \frac{S_\nu}{A_{\text{tot}}}, ~~ \ V_2(\nu) = \frac{C_{>\nu} - C_{<\nu}}{A_{\text{tot}}}.
\end{align}

\section{Specifications of  WPH models}
\label{AppendixFinalModels}


\subsection{WPH Statistics}
\label{AppWPHStats}

We build the specific WPH statistics used in this paper for Fisher analysis and statistical syntheses from two types of WPH moments: $\mathcal{S}$ moments describing a single wavelet frequency band and $\mathcal{C}$ moments describing two. Both are built from the basic WPH moments discussed in Sec.~\ref{PartWPH_Concept} and defined by Eq.~\eqref{EqDefWPH_Cov}, which we repeat here for reference:
\begin{equation}
\tag{\ref{EqDefWPH_Cov}}
\displaystyle{
C_{\vec{\xi}_1,p_1,\vec{\xi}_2,p_2}(\vec{\tau}) = \text{Cov} \left(\left[\rho * \psi_{\vec \xi_1} (\vec{x}) \right]^{p_1},\left[\rho * \psi_{\vec \xi_2} (\vec{x} + \vec{\tau}) \right]^{p_2} \right),
}
\end{equation}
where $\vec{\xi}_i$ are wavelet frequencies, $\left[ \; \right]^{p}$ denotes the $p$th phase harmonic as defined in Eq.~\eqref{EqDefPhaseHarmonic}, and $\vec{\tau}$ is a spatial shift.

The moments used for our statistics are:
\begin{align}
    \label{AppeqdefSsingle}
    \mathcal{S}^{(p_1,p_2)}_{\vec \xi_1}(\vec{\tau}) 
    &= \mathcal{C}_{\vec \xi_1,p_1,\vec \xi_1,p_2} (\vec{\tau}),
    \\[6 pt]
    \label{AppeqdefSdouble} 
    \mathcal{C}^{(p_1,p_2)}_{\vec \xi_1,\vec \xi_2} (\vec \tau) 
    &= \mathcal{C}_{\vec \xi_1,p_1,\vec \xi_2,p_2}(\vec\tau),
    \\[6 pt]
    \label{AppEqPhaseCoupling}
    \mathcal{C}^\text{phase}_{\vec \xi_1,\vec \xi_2}(\vec{\tau}) 
    &= \mathcal{C}_{\vec \xi_1,1,\vec \xi_2,\xi_1/\xi_2}(\vec\tau).
\end{align}
Specifically, we use moments $\mathcal{S}^{(1,1)}$, $\mathcal{S}^{(0,0)}$, $\mathcal{S}^{(0,1)}$, $\mathcal{C}^{(0,0)}$, $\mathcal{C}^{(0,1)}$, and $\mathcal{C}^\text{phase}$, and we restrict to $\xi_2 \le \xi_1$. The physical significance and motivation for using these moments is discussed in Sec.~\ref{PartDiscussionVariousTerms}.

We also define WPH moments $\mathcal{S}_\text{isopar}$ and $\mathcal{C}_\text{isopar}$ that are invariant under rotation and parity, based on the invariant WPH moments discussed in Sec.~\ref{PartSymmetries},
\begin{equation}
\tag{\ref{EqPhaseCoupling_Iso}}
\displaystyle{
{\mathcal{C}}^{\text{isopar}}_{j_1,p_1,j_2,p_2,\delta \ell}(\vec{\tau}) = 
\left \langle 
\mathcal{C}_{j_1,\ell_1,p_1,j_2,\ell_2, p_2} (\vec{\tau})
\right \rangle_{  |\ell_2-\ell_1| = \delta\ell} },
\end{equation}
where $\delta\ell \ge 0$ is an absolute angle, $\langle \; \rangle $ denotes an angular average (over $\ell_1$ and $\ell_2$), and the moment $\mathcal{C}_{j_1,\ell_1,p_1,j_2,\ell_2, p_2}$ refers to the standard WPH moment $\mathcal{C}_{\vec\xi_1,p_1,\vec\xi_2,p_2}$ of Eq.~\eqref{EqDefWPH_Cov} with the $\vec\xi$ and $(j,\ell)$ indices related by Eq.~\eqref{EqjellTok} in the usual way. Specifically, the invariant moments for our statistics are:
\begin{align}
    \label{AppeqdefSsingleIsopar}
    \mathcal{S}^{(p_1,p_2)}_{\text{isopar},j_1}(\vec{\tau}) 
    &= \mathcal{C}^{\text{isopar}}_{j_1,p_1,j_1,p_2,0}(\vec{\tau}), 
    \\[6 pt]
    \label{AppeqdefSdoubleIsopar} 
    \mathcal{C}^{(p_1,p_2)}_{\text{isopar},j_1,j_2,\delta\ell} (\vec \tau) 
    &= \mathcal{C}^\text{isopar}_{j_1,p_1,j_2,p_2,\delta\ell}(\vec\tau),
    \\[6 pt]
    \label{AppEqPhaseCouplingIsopar}
    \mathcal{C}^\text{phase}_{\text{isopar},j_1,j_2,\delta\ell}(\vec{\tau}) 
    &= \mathcal{C}^\text{isopar}_{j_1,1,j_2,2^{j_2 - j_1},\delta\ell}(\vec\tau),
\end{align}
with the restriction $j_1 \le j_2$.

To construct these statistics, we consider only a discrete set $\{\tau_{n,\alpha}\}$ of spatial translations labeled by an integer $n$ and an angle $\alpha$. Translation $\tau_{n,\alpha}$ is defined with respect to the wavelet $\psi_{j,\ell}$ of largest characteristic wavelength appearing in Eq.~\eqref{EqDefWPH_Cov}, and is oriented at an angle $\alpha$ relative to the direction of oscillation of this wavelet:
\begin{equation}
\label{eqDefTauNTheta}
\tau_{n,\alpha} = n 2^j \vec{e}_{(2\pi \ell/L) + \alpha},
\end{equation}
where $\vec{e}$ is the unit vector in the specified direction.
We use integer values $n$ ranging from 0 to $\Delta_n$, with various choices of $\Delta_n \le 5$ depending on the specific type of moment and the values of $j_1,j_2$. For the invariant (isopar) statistics, we use $\alpha = 0$. Otherwise we restrict $\alpha$ to integer multiples of $\pi/4$.
Note that these translations are redundant when larger than half of the size of the fields (128 pixels in this paper).


\subsection{Models for Fisher analysis}
\label{AppFisherWPHModel}

\begin{table}[t]
\centering
\begin{tabular}{| c | c | c | c | c | c | }
\hline
\xrowht{17.5pt}
Model & I & II & III & IV & V  \\
\hline
\xrowht{17.5pt}
Moments & $\mathcal{S}^{(1,1)}_\text{isopar}$ & + $\mathcal{S}^{(0,0)}_\text{isopar}$ & + $\mathcal{S}^{(0,1)}_\text{isopar}$ & + $\mathcal{C}^{(0,0)}_\text{isopar}$, $\mathcal{C}^{(0,1)}_\text{isopar}$ & + $\mathcal{C}^{\text{phase}}_\text{isopar}$  \\
\hline
\xrowht{17.5pt}
Total size & 32 & 64 & 96 & 252 & 327 \\
\hline
\end{tabular}
\caption{The WPH moments included in the nested models of Section~\ref{PartDiscussionVariousTerms} to perform Fisher analysis.}
\vskip 1.5\baselineskip
\label{TableModelsFisher}     
\end{table}

We present in this section the WPH statistics used for the Fisher analysis of Secs.~\ref{PartCosmoInfo} and~\ref{PartDiscussionVariousTerms}. These statistics have integer $j$ values from 0 and 7, corresponding to wavelengths from 2 to 256 pixels. Thus, all the scales of the matter density fields are characterized with WPH moments. These WPH statistics are also invariant under rotation and parity, and characterize couplings between all wavelet bands, with $\Delta j = 7$. They are defined as follows:
\begin{itemize}
\item For $\mathcal{S}^{(1,1)}_\text{isopar}$, $\mathcal{S}^{(0,0)}_\text{isopar}$, and $\mathcal{S}^{(0,1)}_\text{isopar}$ moments, we consider all $j_1$ values from 0 to 7. For each $j_1$ value, we consider all possible translations $\tau_{n,\alpha}$ with $\alpha = 0$ and $0 \leq n \leq \Delta_n(j_1)$, where %
$[\Delta_n(0), \ldots, \Delta_n(7)] = [5, 5, 5, 5, 3, 1, 0, 0]$. 
\item For $\mathcal{C}^{(0,0)}_\text{isopar}$, $\mathcal{C}^{(0,1)}_\text{isopar}$, and $\mathcal{C}^{\text{phase}}_\text{isopar}$, we consider all $(j_1,j_2)$ pairs satisfying $ 0\leq j_1 \leq j_2 \leq 7 $.
For $j_2=7$, we take only $\delta \ell=0$.
For $j_2<7$, we take $2\pi \delta \ell/L\in\{0, \pi/4, \pi/2\}$ when  $j_1\in\{0,1\}$, and we take $2\pi \delta \ell/L\in\{0,\pi/2\}$ otherwise.
Finally, we consider translations $\tau_{n,\alpha}$ with $n\in\{0, 1\}$ if $0\leq j_1 = j_2 \leq 5$ and $\delta \ell=0$, and no translations otherwise (i.e., we keep $n=0$ only).
Note that $\mathcal{C}$ moments with $j_1=j_2$ and $\delta\ell=0$ are equal to $\mathcal{S}$ moments and are therefore omitted to avoid double counting.
\end{itemize}
These WPH statistics contain a total of 327 WPH moments. 
In Sec.~\ref{PartDiscussionVariousTerms}, we divide these WPH statistics into a set of five nested models, model~I to model~V, to assess how the Fisher information about cosmological parameters increases as WPH moments with different properties and physical significance are included. These models are summarized in Table~\ref{TableModelsFisher}.
%
%
%

\begin{table}[b]
\centering
\begin{tabular}{| c | c | c | c | c | c | }
\hline
\xrowht{17.5pt}
Model & I & II & III & IV & V  \\
\hline
\xrowht{17.5pt}
Moments & $\mathcal{S}^{(1,1)}$ & + $\mathcal{S}^{(0,0)}$ & + $\mathcal{S}^{(0,1)}$ & + $\mathcal{C}^{(0,0)}$, $\mathcal{C}^{(0,1)}$ & + $L_{j, p}$,  $\mathcal{C}^{\text{phase}}$  \\
\hline
\xrowht{17.5pt}
Total size & 677 & 1349 & 2021 & 5412 & 6676 \\
\hline
\end{tabular}
\caption{The WPH moments included in the nested models of Section~\ref{PartDiscussionVariousTerms} to perform statistical syntheses.}
\label{TableModelsSynthesis}     
\end{table}

\begin{figure*}[t]
\begin{subfigure}{.32\textwidth}
\begin{center}
\includegraphics[width = 0.99\textwidth]{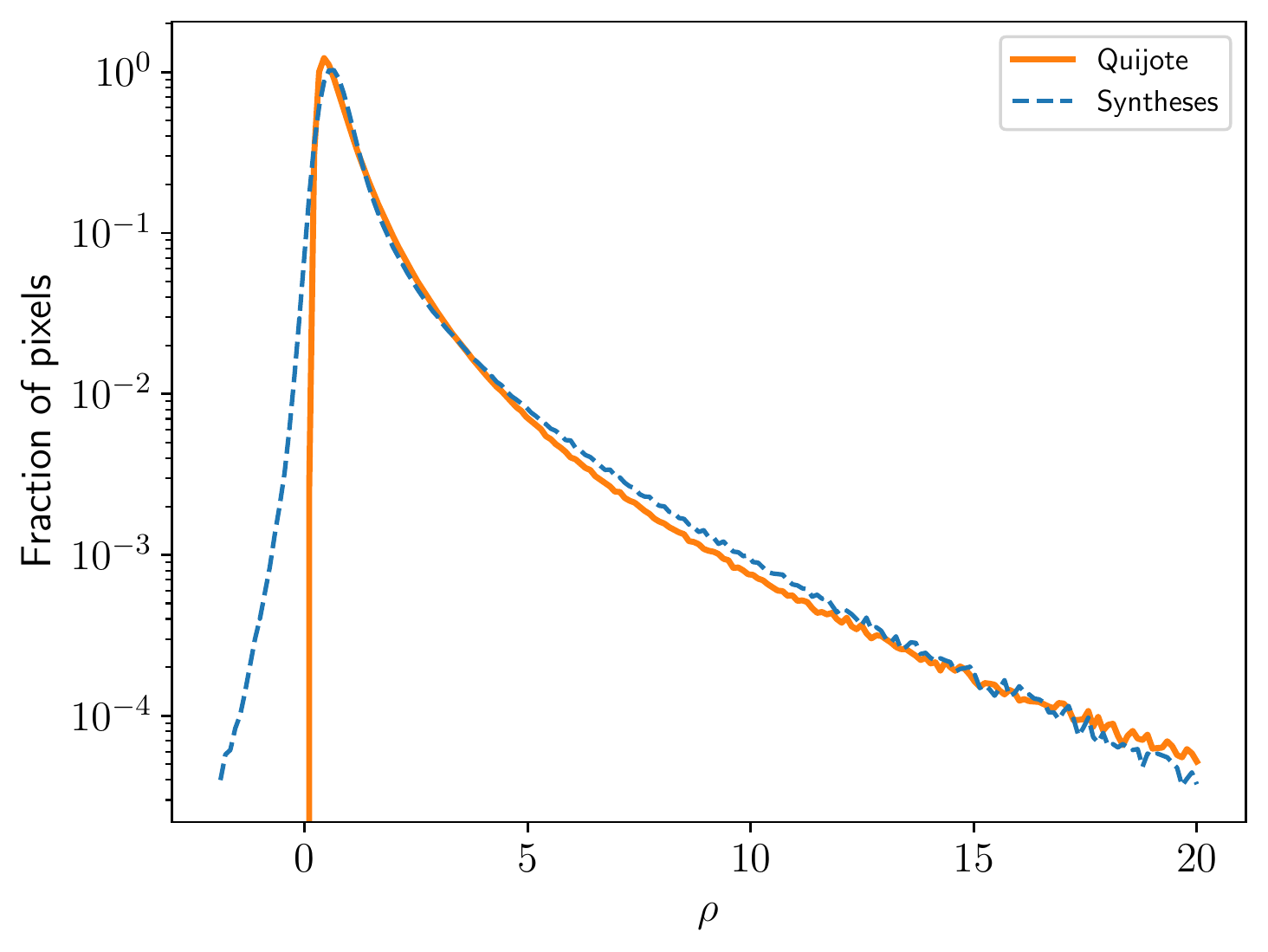}
\vspace{-0.15cm}
\end{center}
\end{subfigure}
\begin{subfigure}{.32\textwidth}
\begin{center}
\includegraphics[width = 0.99\textwidth]{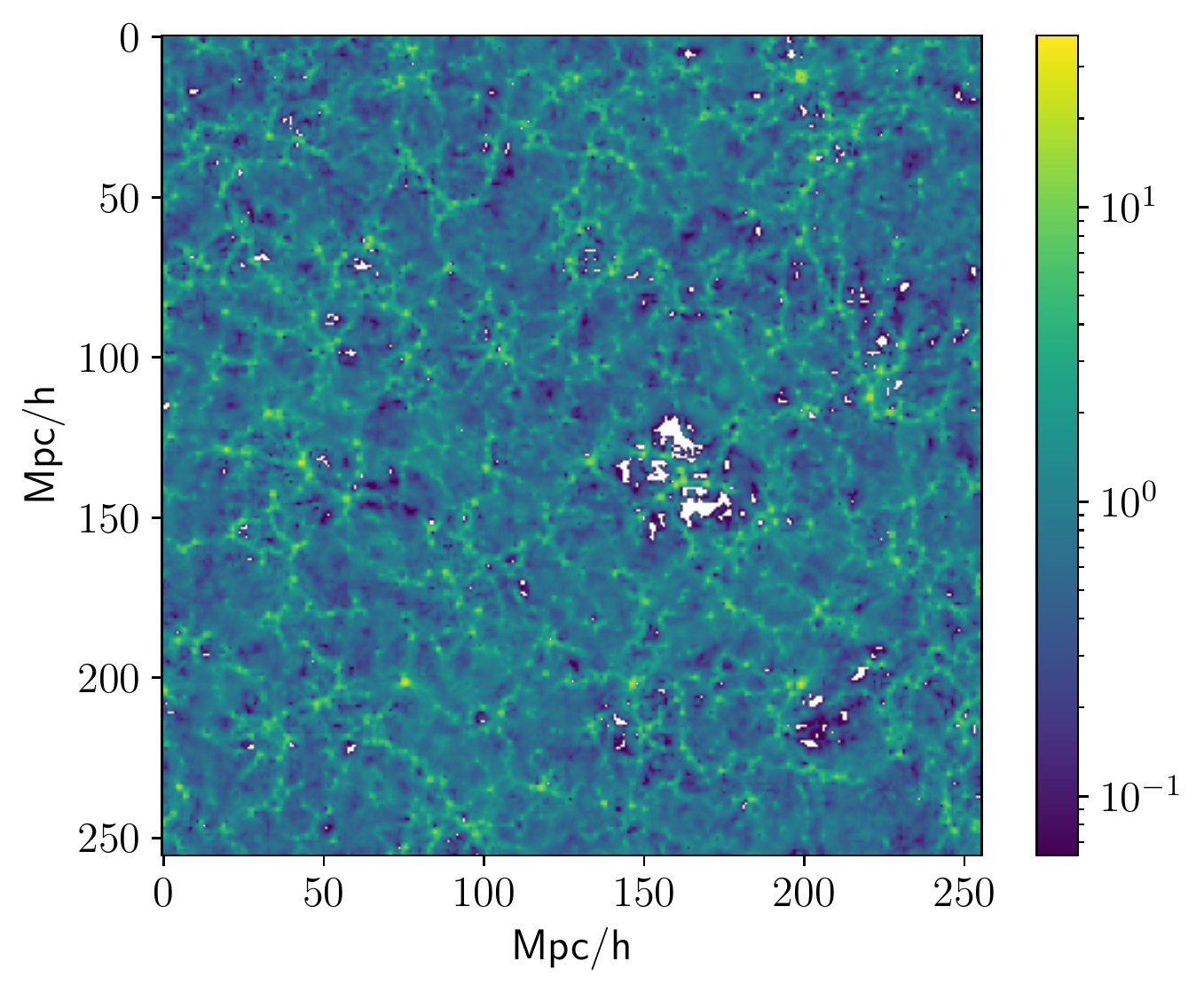}
\vspace{-0.15cm}
\end{center}
\end{subfigure}
\begin{subfigure}{.32\textwidth}
\begin{center}
\includegraphics[width = 0.99\textwidth]{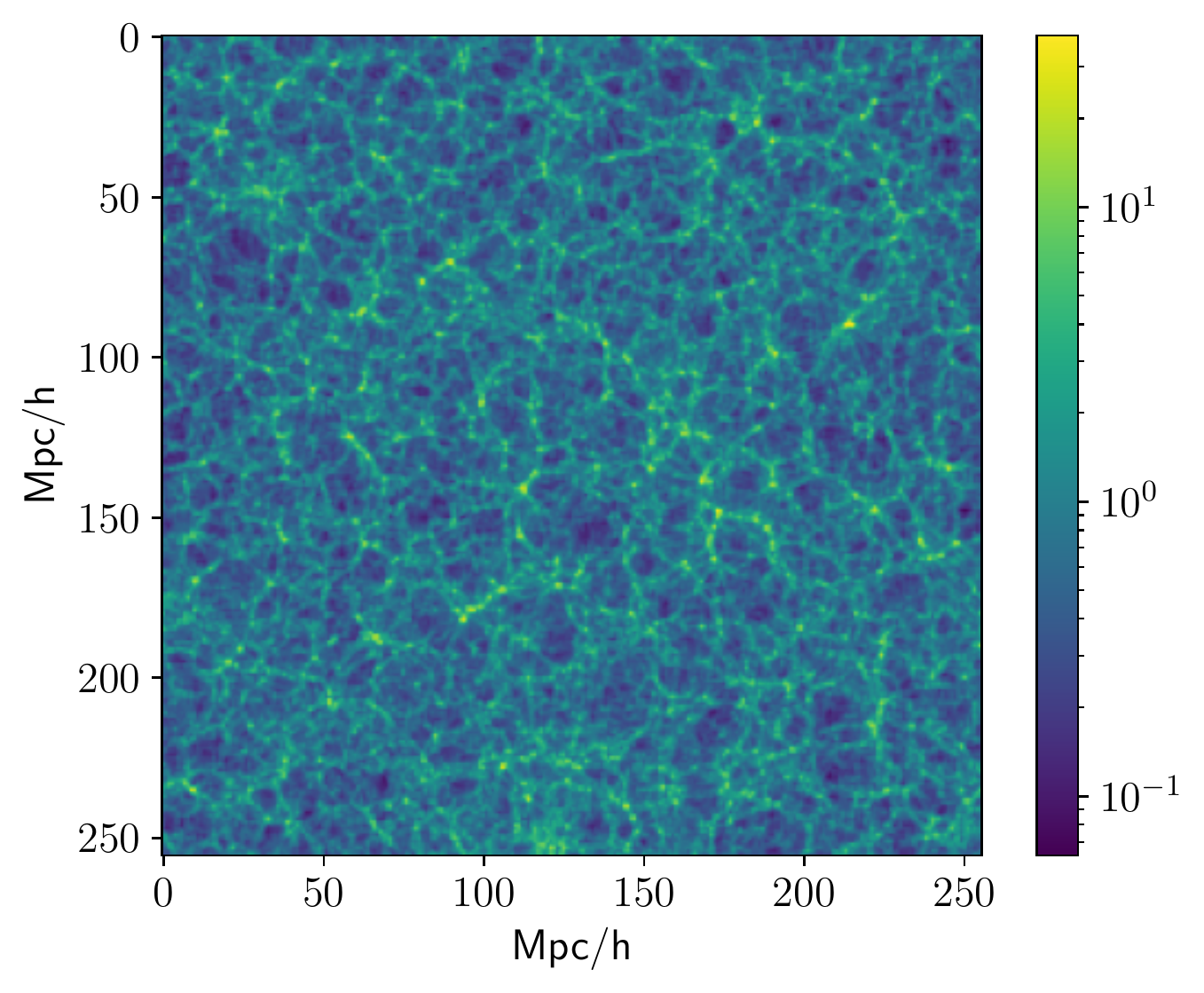}
\vspace{-0.15cm}
\end{center}
\end{subfigure}
 \caption{Left: histogram of the Quijote density field $\rho$ and of the syntheses of $\rho$. The syntheses are not strictly positive, contrary to the initial maps. Middle: example of a syntheses of $\rho$ shown in logarithmic scales. The white pixels correspond to negative value of the field. Right: example of a Quijote simulation $\rho$.}
\label{FigSynthExpField}
\end{figure*}

\subsection{Models for statistical syntheses}
\label{AppSynthesesWPHModel}

This section presents the WPH statistics used to perform syntheses in Sec.~\ref{PartSyntheses} and~\ref{PartDiscussionVariousTerms}. The WPH statistics used in the syntheses of Sec.~\ref{PartSyntheses} have integer $j$ values from 0 to 5, corresponding to wavelengths up to 64 pixels. They thus do not characterize in terms of WPH moments the largest scales of the matter density fields, but describe couplings between all scales up to $j=5$, with $\Delta j = 5$. The moments used for these statistics are not required to be invariant under rotation and parity, contrary to those used for Fisher analysis. They are defined as follows:
\begin{itemize}
\item For $\mathcal{S}^{(1,1)}$, $\mathcal{S}^{(0,0)}$, and $\mathcal{S}^{(0,1)}$ moments, we consider all $j_1$ values from 0 to 5 and $\ell_1$ values from $0$ to $15$. For each $j_1$ value, we consider for $\mathcal{S}^{(1,1)}$ and $\mathcal{S}^{(0,0)}$ all possible translations $\tau_{n,\alpha}$ with $0 \leq n \leq \Delta_n = 2$ and $\alpha \in \{-\pi/4, 0, \pi/4, \pi/2\}$. No translations are applied to $\mathcal{S}^{(0,1)}$.
\item For $\mathcal{C}^{(0,0)}$, $\mathcal{C}^{(0,1)}$, and $\mathcal{C}^{\text{phase}}$, we consider all $(j_1,j_2)$ pairs satisfying $ 0\leq j_1 \leq j_2 \leq 5 $, and all $\ell_1$ values $0 \leq \ell_1 < L = 16$. We take $\delta \ell \equiv \ell_2 - \ell_1 = 0$ for $\mathcal{C}^{\text{phase}}$, and $2\pi |\delta \ell|/L\in\{0, \pi/8, \pi/4, 3\pi/8, \pi/2\}$  for  $\mathcal{C}^{(0,0)}$ and $\mathcal{C}^{(0,1)}$. For $\mathcal{C}^{(0,1)}$ and $\mathcal{C}^{\text{phase}}$, when $\delta \ell = 0$, we apply all possible translations $\tau_{n,\alpha}$ with $0 \leq n \leq \Delta_n = 2$ and $\alpha \in \{-\pi/4, 0, \pi/4, \pi/2\}$. No translations are applied (i.e., we take $\Delta_n=0$) for $\mathcal{C}^{(0,1)}$ when $\delta \ell \neq 0$, or for $\mathcal{C}^{(0,0)}$ with any $\delta\ell$.
$\mathcal{C}$ moments with $j_1=j_2$ and $\delta\ell=0$ are omitted because they are equal to $\mathcal{S}$ moments.
\end{itemize}

To complete these WPH statistics and better constrain the scales that are not probed by WPH moments as well as the probability density function, we also consider convolutions of the field $\rho$ with a family of low-pass filters $\varphi_{j}(\vec{x})$ called \textit{scaling functions}, defined in  Appendix~\ref{AppendixScalingFunction}. We therefore added the following scaling moments $L_{j, p}$ to the WPH moments:
\begin{align}
\label{EqLjp}
L_{j,0} &= \text{Cov} \left[\lvert\rho * \varphi_{j} \rvert,\lvert\rho * \varphi_j \rvert \right],\\
L_{j, p} &= \text{Cov} \left[\left(\rho * \varphi_{j} \right)^{p},\left(\rho * \varphi_j \right)^{p} \right]~~~(\text{for}~p>0),
\end{align}
with $j$ between $2$ and $5$ and $p\in \{0,1,2,3\}$, yielding 16 scaling moments. 

This model (which corresponds to model~V of Sec.~\ref{PartSyntheses}) contains overall 6660 WPH moments and 16 scaling moments. In Sec.~\ref{PartDiscussionVariousTerms} we consider syntheses using nested models, as summarized in Table~\ref{TableModelsSynthesis}. 

\section{Limitations of the model and direct syntheses of the field $\rho$}
\label{AppLimitations}

We applied the method presented in section \ref{PartSyntheses} to directly generate syntheses of the raw density field $\rho$ (instead of its logarithmic value). However, this field $\rho$ is strictly positive, with a wide dynamic range. The resulting syntheses did not reproduce the sharp constraint $\rho > 0$. An example of these syntheses and their histogram is provided in Fig.\ref{FigSynthExpField}. This figure, however, shows that the filamentary structure of the LSS density field is nevertheless recovered.

Another possibility to obtain strictly positive syntheses of the density field $\rho$ is to take the exponential of the syntheses of $\log \rho$ presented in section \ref{PartSyntheses}. The histograms and the Minkowski functionals of the resulting syntheses reproduce the statistics of the simulations with an accuracy similar to the results presented in \ref{sec:resultSynth}. However, the other statistics are not well recovered. For example, the mean and the standard deviation of the power spectrum of the syntheses differ from those of the simulations by respectively $30\%$ and $50\%$ for most of the spatial frequency $k$.

\vspace{2cm}


%

\end{document}